\documentclass{emulateapj}
\usepackage{amsmath}
\usepackage{natbib}
\newcommand{\gps}{\ensuremath{g_{\rm P1}}}
\newcommand{\rps}{\ensuremath{r_{\rm P1}}}
\newcommand{\ips}{\ensuremath{i_{\rm P1}}}
\newcommand{\zps}{\ensuremath{z_{\rm P1}}}
\newcommand{\yps}{\ensuremath{y_{\rm P1}}}

\newcommand{\PS}{\protect \hbox {Pan-STARRS1}}
\newcommand{\degree}{\ensuremath{^\circ}}
\newcommand{\dfplot}[1]{\plotone{#1}}

\newcommand{\m}{\ensuremath{\vec{m}}}
\renewcommand{\mp}{\ensuremath{\left\{ \m \right\}}}
\renewcommand{\a}{\ensuremath{\vec{\alpha}}}

\newcommand{\noprint}[1]{}
\newcommand{\figsetstart}{{\bf Fig. Set} }
\newcommand{\figsetend}{}
\newcommand{\figsetgrpstart}{}
\newcommand{\figsetgrpend}{}
\newcommand{\figsetnum}[1]{{\bf #1.}}
\newcommand{\figsettitle}[1]{ {\bf #1} }
\newcommand{\figsetgrpnum}[1]{\noprint{#1}}
\newcommand{\figsetgrptitle}[1]{\noprint{#1}}
\newcommand{\figsetplot}[1]{\noprint{#1}}
\newcommand{\figsetgrpnote}[1]{\noprint{#1}}

\shorttitle{Distances to Molecular Clouds from PS1}
\shortauthors{E. F. Schlafly et al.}
\begin{document}
\title{A Large Catalog of Accurate Distances to Molecular Clouds from PS1 Photometry}
\author{
E. F. Schlafly,\altaffilmark{1}
G. Green,\altaffilmark{2}
D. P. Finkbeiner,\altaffilmark{2,3}
H.-W. Rix,\altaffilmark{1}
E. F. Bell,\altaffilmark{4}
W. S. Burgett,\altaffilmark{5}
K. C. Chambers,\altaffilmark{5}
P. W. Draper,\altaffilmark{6}
K. W. Hodapp,\altaffilmark{5}
N. Kaiser,\altaffilmark{5}
E. A. Magnier,\altaffilmark{5}
N. F. Martin,\altaffilmark{7,1}
N. Metcalfe,\altaffilmark{6}
P. A. Price,\altaffilmark{8}
J. L. Tonry\altaffilmark{5}
}

\altaffiltext{1}{Max Planck Institute for Astronomy, K\"{o}nigstuhl 17, D-69117 Heidelberg, Germany}
\altaffiltext{2}{Harvard-Smithsonian Center for Astrophysics, 60 Garden Street, Cambridge, MA 02138}
\altaffiltext{3}{Department of Physics, Harvard University, 17 Oxford Street, Cambridge MA 02138}
\altaffiltext{4}{Department of Astronomy, University of Michigan, 500 Church St., Ann Arbor, MI 48109}
\altaffiltext{5}{Institute for Astronomy, University of Hawaii, 2680 Woodlawn Drive, Honolulu HI 96822}
\altaffiltext{6}{Department of Physics, Durham University, South Road, Durham DH1 3LE, UK} 
\altaffiltext{7}{Observatoire Astronomique de Strasbourg, CNRS, UMR 7550, 11 rue de l'Universit\'{e}, F-67000 Strasbourg, France}
\altaffiltext{8}{Department of Astrophysical Sciences, Princeton University, Princeton, NJ 08544, USA} 

\begin{abstract}
Distance measurements to molecular clouds are important, but are often made separately for each cloud of interest, employing very different different data and techniques.  We present a large, homogeneous catalog of distances to molecular clouds, most of which are of unprecedented accuracy.  We determine distances using optical photometry of stars along lines of sight toward these clouds, obtained from PanSTARRS-1.  We simultaneously infer the reddenings and distances to these stars, tracking the full probability distribution function using a technique presented in \citet{Green:2014}.  We fit these star-by-star measurements using a simple dust screen model to find the distance to each cloud.  We thus estimate the distances to almost all of the clouds in the \citet{Magnani:1985} catalog, as well as many other well-studied clouds, including Orion, Perseus, Taurus, Cepheus, Polaris, California, and Monoceros R2, avoiding only the inner Galaxy.  Typical statistical uncertainties in the distances are 5\%, though the systematic uncertainty stemming from the quality of our stellar models is about 10\%.  The resulting catalog is the largest catalog of accurate, directly-measured distances to molecular clouds.  Our distance estimates are generally consistent with available distance estimates from the literature, though in some cases the literature estimates are off by a factor of more than two.
\end{abstract}

\keywords{ISM: dust, extinction --- ISM: clouds}

\section{Introduction}
\label{sec:intro}

Molecular clouds are the site of star formation, where all stars are born \citep{Blitz:1999}.  The study of molecular clouds then informs critical elements of astrophysics, like the initial mass function of stars and the build-up of galaxies.  Intense study has focused on the Milky Way's molecular clouds, the nearest and most accessible sites of star formation.  The distances to these clouds are fundamental to deriving their basic physical parameters---like mass and size---from observations.  But estimating the distance to molecular gas is difficult, and a number of different techniques have been explored and applied, often only to individual clouds of interest.

These techniques are varied.  A common method is to estimate cloud distances kinematically.  In this technique a cloud's recessional velocity is measured by the Doppler shift of its spectral lines and converted to a distance by assuming that the cloud follows the Galactic rotation curve.  This technique is widely applicable and has been used to estimate the distances to large numbers of molecular clouds (e.g., \citet{RomanDuval:2009}), but it is problematic in the presence of peculiar velocities and non-circular motions.  A second method is to find the distance to objects associated with a cloud and to place the cloud at the same distance; for instance, many clouds have formed young OB associations of stars for which distances can be estimated.  

A third method is to estimate a cloud's distance from its reddening and absorption of starlight.  Light passing through molecular clouds is extinguished by dust and gas; in particular, optical and infrared light is reddened by dust.  This allows stars in the foreground of the cloud to be distinguished from stars in its background.  By finding the distances to these stars, the distance to the cloud can be determined.  Recently \citet{Lallement:2014} and \citet{Vergely:2010} have mapped the 3D distribution of the ISM in the solar neighborhood using this basic technique.  A systematic study  \citep{Lombardi:2001, Lombardi:2011, Lada:2009} using data from 2MASS has led to precise distance estimates for a number of clouds by counting the number of unextinguished foreground stars toward large molecular clouds and comparing with predictions for the distribution of stars from the Besan\c{c}on Galactic model \citep{Robin:2003}.

We have developed a related technique: we simultaneously infer the distance and reddening to stars from their \PS\ \citep[PS1]{PS1_system} photometry and bracket clouds between foreground unreddened stars and background reddened stars.  The use of only \PS\ photometry gives us access to three-quarters of the sky and hundreds of millions of stars, but has the disadvantage that the distances and reddenings we infer have strongly covariant uncertainties.  We track the full probability distribution function of distance and reddening to each star, and model the results as produced by a screen of dust associated with the cloud, with an angular distribution given by the Planck dust map \citep{Planck:2011}.  We then perform an MCMC sampling to determine the range of probable distances to the cloud.

This paper is part of an ongoing effort to study the dust using \PS\ photometry.  The basic method is presented in \citet{Green:2014}, while E. Schlafly et al. (2014, in preparation) demonstrates that the technique closely reproduces the widely-used reddening map of \citet[SFD]{Schlegel:1998}.  This work serves additionally to demonstrate the 3D power of the method, recovering the distances to the Galaxy's molecular clouds.

We measure the distances to many well-studied molecular clouds in the Galaxy: Orion, $\lambda$ Orionis, Taurus, Perseus, California, Ursa Major, the Polaris Flare, the Cepheus Flare, Lacerta, Pegasus, Hercules, Camelopardis, Ophiuchus, and Monoceros R2.  We additionally estimate the distances to most of the clouds of the \citet{Magnani:1985} catalog of high Galactic latitude molecular clouds, though in some cases the cloud does not fall within the PS1 footprint.  Our distances are often consistent with, but more precise than, other available distance estimates, though we find that occasionally the literature distance estimates are off by as much as a factor of two.  In this work we avoid clouds in the inner Galaxy.  In principle we could apply this technique there as well, but these clouds require more sophisticated modeling of the potentially many molecular clouds on each line of sight through the disk.  We accordingly defer analysis of the inner Galaxy to later work.

We describe in \textsection\ref{sec:ps1} the \PS\ survey, which provides the optical photometry on which this work is based.  In \textsection\ref{sec:method} we describe our method for determining the distances to the dust clouds.  In \textsection\ref{sec:results}, we apply our technique to sight lines through molecular clouds in the \PS\ footprint, and present a catalog of cloud distances.  In \textsection\ref{sec:sysunc} and \textsection\ref{sec:discussion}, we discuss the systematic uncertainties in the method and the implications of the results in light of the literature.  Finally, we conclude in \textsection\ref{sec:conclusion}.

\section{The PanSTARRS-1 System and Surveys}
\label{sec:ps1}

The \PS\ survey provides homogeneous, five-filter, optical and near-infrared photometry of the entire sky north of $\delta = -30\degree$, making it well suited to this analysis.  The \PS\ system is situated on Haleakala \citep{PS1_system}, and regularly delivers arcsecond seeing.  The 1.8~m telescope has a $3\degree$ field of view outfitted with the 1.4 billion pixel GPC1 camera \citep{PS1_optics, PS1_GPCA, PS1_GPCB}.  Images from the telescope are processed nightly by the Image Processing Pipeline, which automatically corrects bias and dark signatures, flattens images, and performs astrometry and photometry \citep{PS1_IPP, PS1_photometry, PS1_astrometry}.  

This analysis relies on the PS1 $3\pi$ survey (K. Chambers et al., in preparation), in which each part of the sky is observed 4 times each year in each of five filters, denoted $griz\yps$.  Typical $5\sigma$ single-epoch depths are 22.0, 22.0, 21.9, 21.0, 19.8 in $griz\yps$, with stacked data going about 1.1~mag deeper \citet{Metcalfe:2013}.  These filters are close analogs of the filters used in the SDSS survey, and differ primarily in that the \zps\ filter is cut off at 920~nm and that the SDSS $u$ filter is traded for a \yps\ filter which covers 920~nm to 1030~nm \citep{PS_lasercal}.  The photometric calibration is based on \citet{JTphoto} and \citet{Schlafly:2012}, which respectively provide the absolute and relative photometric calibration of the survey with better than $1\%$ accuracy.  We use data from the first major uniform processing of the PS1 data, dubbed Processing Version 1, which primarily includes images taken between May 2010 and March 2013.  

We use PS1 single-epoch data, and average together multiple observations of the same object.  The PS1 single-epoch data at present provides the most accurate photometry for bright stars; in future work we will adopt the stacked data to reach fainter stars and larger distances.  We only use objects which have been detected in the \gps\ filter and at least three of the four $riz\yps$ filters, to restrict ourselves to objects for which our distance estimates are most accurate.  We exclude galaxies from the analysis by requiring that the aperture magnitude of the object be less than 0.1 mag brighter than the PSF magnitude in at least three bands.  This is a relaxed cut and was chosen to produce a relatively clean stellar locus at high Galactic latitudes.  Finally, analysis of repeated detections of the same objects indicates that the \PS\ pipeline somewhat underestimates its photometric uncertainties; we adopt modified uncertainties by inflating the pipeline uncertainties by 20\% and adding 15 mmag in quadrature with them.

\section{Method}
\label{sec:method}

We find the distances to dust clouds in a two-step process.  First, we determine the reddenings and distances to individual stars in a certain direction using their \PS\ photometry.  Second, we combine this information to determine the reddening as a function of distance, $E(D)$ (or \emph{reddening profile}), adopting a simple dust screen model for $E(D)$.  This model consists of uniform nearby screen of reddening (assumed to be more nearby than all the stars observed by PS1) and a thin screen of dust at the distance to the cloud.  The angular structure of the cloud screen is adopted from the Planck dust map \citep{Planck:2011}.  The basic idea is to bracket the cloud distance by the distances to unreddened foreground stars and reddened background stars, considering the full covariance of the uncertainties in reddening and distance to each of the stars.

Specifically, we adopt the technique of \citet{Green:2014} to infer for each star the function $p(E,D)$, which describes the full probability distribution function of the star's reddening $E$ and distance $D$, subject to its \PS\ photometry.  This technique uses a set of stellar models giving the intrinsic colors of stars as a function of their absolute magnitude and metallicity \citep{Ivezic:2008}.  It additionally folds in prior expectations about the distribution of stars in space \citep{Juric:2008}, metallicity \citep{Ivezic:2008} and luminosity \citep{Bressan:2012}.  We use throughout a fixed $R_V = 3.1$ reddening law from \citet{Fitzpatrick:1999}, adapted to the PS1 bands according to \citet{Schlafly:2011}.  As described in \citet{Green:2014}, we determine the range of probable absolute magnitudes, distances, metallicities, and reddenings for each star by comparing its observed photometry with that expected from the models, and Markov Chain Monte Carlo (MCMC) sampling the distribution.  By marginalizing out absolute magnitude and metallicity, we obtain $p(E,D)$, completely describing the reddening and distance to the star.  As shown in \cite{Green:2014}, if the reddening profile $E_{\a}(D)$ is parameterized by $\a$, then the probability distribution function $p(\a \mid \mp)$ is given by
\begin{equation}
\label{eq:palpha}
p(\a \mid \mp) \propto p(\a) \prod_i \int dD p(E(D), D \mid \m_i)
\end{equation}
where $\mp$ gives the photometry for all objects along a line of sight, $i$ indexes over stars, and $p(E,D \mid \m_i)$ is, up to a normalizing constant, the probability distribution function of reddening and distance for star $i$, when a flat prior is adopted on $E$.  The parameters $\a$ are then determined ultimately by how they affect the integral through $p(E, D)$ along the line $E(D)$.

\begin{figure}[h]
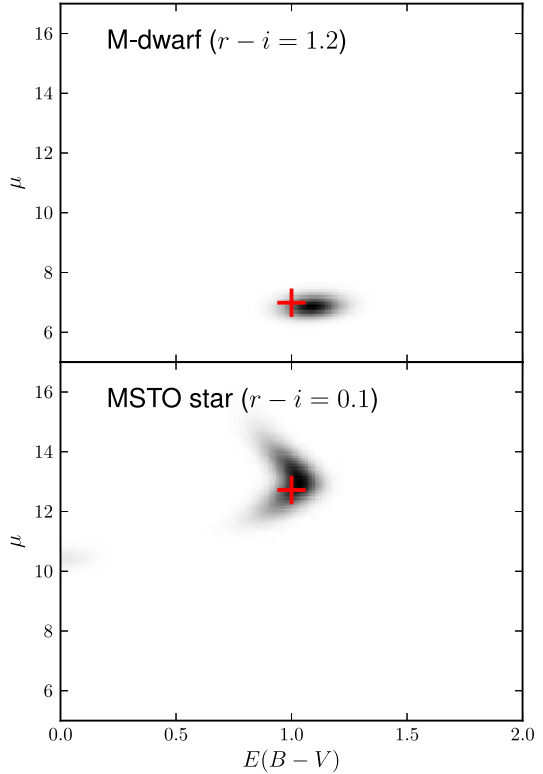

\dfplot{pad-redblue}
\figcaption{
\label{fig:pad} Distance and reddening estimates for two stars: a red M-dwarf with intrinsic $\rps-\ips = 1.2$ and a blue main-sequence turn-off (MSTO) star with intrinsic $\rps-\ips = 0.1$.  We simulate PS1 photometry for stars with these intrinsic colors, a reddening $E(B-V) = 1$, and distances of 250~pc and 3.5~kpc, respectively.  Given this photometry, we obtain the probability of a star having a particular reddening and distance modulus shown by the grayscale.  The true reddening and distance of these stars (red crosses) are near the peak of the grayscales, suggesting a successful fit; see \citet{Green:2014} for detailed statistical tests of the method.  The shape of the stellar locus in the PS1 colors leads to dramatically different probability distributions for red and blue stars.  For an M-dwarf (top), the distance and reddening uncertainties are less correlated and the distances are better constrained than for a MSTO star (bottom).  The shape of the probability distribution for the MSTO star is characteristic: the curve in the diagram is associated with the MSTO in our stellar models, and the small amount of probability that the star be unreddened and at $\mu = 10$ corresponds to the possibility that the reddened MSTO star is in fact an unreddened early M-dwarf.
}
\end{figure}

The work of \citet{Green:2014} details the shape of $p(E,D \mid \m)$ for different types of stars.  Because these surfaces underlie the work presented here, we summarize the discussion there briefly.  Two typical surfaces $p(E,D)$ are shown in Figure~\ref{fig:pad}: one for a blueish main-sequence turn-off star and another for a red M-dwarf.  Because distances are determined from the difference between observed magnitudes $m$ and intrinsic magnitudes $M$, we express distance in terms of the distance modulus $\mu = m - M = 5 \log(D/10~\mathrm{pc})$.  For blue stars the intrinsic color of the star is largely degenerate with the star's reddening, leading to uncertainty of about 0.2~mag in reddening $E(B-V)$.  The intrinsic color uncertainty leads to a significant distance uncertainty.  This situation is especially problematic near the main-sequence turn-off, where both evolved stars and main-sequence stars have similar colors, leading to an especially broad range of allowed distances.  The shape of $p(E,D)$ in the bottom panel of Figure~\ref{fig:pad} is characteristic: on the main sequence, as the reddening increases, bluer intrinsic colors and hence greater intrinsic luminosities and larger distances are implied, while past the main-sequence turn-off ($\mu \approx 13$ for the particular star shown in the bottom panel of Figure~\ref{fig:pad}), redder intrinsic colors imply greater luminosities and hence distance, leading reddening to decrease with distance.  For M-dwarfs (top panel), the degeneracy between reddening and intrinsic color is much less severe: the M-dwarf locus in $g-r$, $r-i$ color-color space is nearly orthogonal to the reddening vector.  Accordingly, for these stars the reddening and especially distance are more tightly constrained.

The work of \citet{Green:2014} and E. Schlafly et al. (2014, in preparation) consider a generic description for the reddening profile $E_{\a}(D)$ which is piecewise-linear in the distance modulus $\mu = 5\log(D/10~\mathrm{pc})$.  However, many dust clouds of particular interest are reasonably isolated objects that dominate the dust column along their line of sight.  Moreover, these clouds are often thin relative to the distance to the cloud.  Accordingly, this motivates modeling the clouds with a more restrictive parameterization of $E_{\a}(D)$.  Therefore, we take $\a$ to be $(D_c, N, f)$: the distance $D_c$ to the single dust cloud along the line of sight, the optical depth $N$ of the cloud, and the nearby foreground reddening $f$.  This is basically a thin dust screen approximation, with the addition of some foreground reddening $f$.  Because emission-based dust maps can provide high resolution data on the angular distribution of dust in the screen, we model the dust extinction through the cloud as $NC$, where $C$ gives the total dust column through the cloud from Planck \citep{Planck:2011}, and $N$ is a normalization factor for the map.  That is, we adopt the model
\begin{equation}
\label{eq:model}
E_i(D) = 
\begin{cases}
f & \text{if } D < D_C \\
f + N C_{i} & \text{if } D \geq D_C
\end{cases}
\end{equation}
where $i$ indexes over stars, and $C_i$ is the estimated total dust column in the direction of star $i$ from the Planck dust map.  In principle we could impose a strong prior on $N$, demanding that the total extinction $E$ be given by the {\it Planck} maps.  However, we have found that the SFD map can have a scale error of about 10\% that varies from cloud to cloud in the sky \citep{Schlafly:2010, Schlafly:2011}, and we find below that $N$ can vary between 0.5--1.2 in the clouds studied in this work (\textsection\ref{sec:results}).  In general we find $N < 1$, as expected as only a fraction of the total reddening along each line of sight is associated with the cloud.  In general, any foreground reddening, background reddening, or diffuse reddening spread out along the line of sight will reduce $N$.  An additional complication is posed by the different effective resolutions of the stellar-based reddenings and the emission map-based reddenings.  The stellar reddenings sample the dust at nearly perfect resolution, while dust emission maps are blurred by the instrumental point spread function.  For the often clumpy and filamentary clouds considered in this analysis, this may lead to $N \neq 1$ even when the reddening associated with the cloud dominates the total reddening.

Our approach (Equation~\ref{eq:palpha}) requires that we adopt priors on the model parameters \a: the cloud distance, the 2D screen normalization factor, and the nearby foreground reddening.  We adopt a simple flat prior on cloud distance $D_c$ and require that the normalization $N$ satisfy $0.2 < N < 2$.  The normalization prior is rarely informative, and serves primarily to make the fit more robust.  In nearby clouds, the foreground reddening $f$ and cloud-associated reddening can be degenerate, since few stars may be foreground to the cloud.  For this reason, we impose the prior that $f$ is less than 25\% of the median projected 2D reddening toward the stars we consider toward each cloud, essentially requiring that the cloud-associated reddening dominate the total projected reddening.

The importance of the Planck-based dust map in this analysis depends on the angular size of the region used to determine the determine cloud distance.  On most lines of sight through distant clouds, we adopt a radius of 0.2\degree, fitting all of the stars within this distance of each sight line.  In many cases, the angular variation of the dust within this region is small, and our fits proceed similarly when assuming no angular variation of the dust and when assuming angular variation from the SFD dust map or the Planck dust map.  In a very few cases, however, the added information provided by the Planck map allows the fit to produce more consistent results than when assuming no angular variation or the angular variation given by SFD.  We attribute this to the complex angular structure of the dust and its temperature, which is mapped with higher resolution by Planck than by SFD.

Because we are essentially looking to find the location of a step in reddening, we exclude any stars from the analysis which have predicted $C_i < 0.15$~mag $E(B-V)$.  This is approximately the $1\sigma$ uncertainty in our reddening estimates for individual stars.  Apparent steps in reddening much smaller than this can stem from limitations in our stellar models or problems in the photometric calibration (\textsection\ref{sec:sysunc}).

When the dust clouds are within approximately 250~pc, few stars are foreground to the cloud within the 0.2\degree\ beam.  This leads to very uncertain distance estimates.  Accordingly, for lines of sight through nearby clouds, we adopt a much larger 0.7\degree\ radius.  We additionally preselect M-stars toward these clouds by making the following cuts on color and magnitude:
\begin{equation}
\label{eq:brightred}
\gps - \frac{A_g}{A_g - A_r} (\gps - \rps - 1.2) < 20 \\
\end{equation}
\begin{equation}
\label{eq:colred}
\rps - \ips - \frac{A_r - A_i}{A_g - A_r} (\gps - \rps) > 0
\end{equation}
Equation~\ref{eq:brightred} is a cut along the reddening vector that selects only bright blue stars, or stars as faint as $\gps = 20$ for unreddened M dwarfs, which have typical $\gps - \rps = 1.2$.  Equation~\ref{eq:colred} is also along the reddening vector, and selects M dwarfs.  The combined cuts result in a very pure, nearby, reddening-independent sample of M dwarfs.  For nearby clouds, limiting the analysis to nearby M dwarfs is valuable because it prevents the small number of foreground stars from being overwhelmed by the large number of background stars in the analysis.  This leads to our adopting two slightly different techniques in this work: a ``far'' technique, where we use all of the stars within a 0.2\degree\ radius line of sight, and a ``near'' technique, where we use only M-dwarfs within a larger 0.7\degree\ radius line of sight.

We determine cloud distances and their uncertainties by MCMC sampling Equation~\ref{eq:palpha} using the model given in Equation~\ref{eq:model} and the \emph{emcee} package of \citet{Foreman-Mackey:2013}.  We report the resulting distance estimates in terms of the 16th, 50th, and 84th percentile of the distance samples from the MCMC chain.  Our implementation is largely a straightforward, but we note here a few implementation details.

We impose a 20 mmag floor on the \PS\ photometric uncertainties in computing $p(E,D)$ for individual stars.  This error floor is motivated by the fact that we do not expect our library of intrinsic stellar types to be accurate at the 20 mmag level.  For instance, following \citet{Ivezic:2008} we neglect variation in color with metallicity on the main sequence in our stellar models in bands redward of the SDSS $u$ band, though we expect that the \PS\ \gps\ band photometry is affected by metallicity at the few hundredth of a magnitude level. 

Our individual star reddening estimates are obtained by comparing observed photometry to model photometry.  Our model photometry is however best suited only for observations of old, main-sequence stars; it includes no objects blueward of the typical main-sequence turn-off for halo stars (e.g., white dwarfs, young high-mass stars, quasars), and its treatment of subgiant, red giant, and asymptotic giant branch stars is rudimentary.  To mitigate this, we ignore in the analysis any stars that have $\chi^2$ for the best fit stellar model more than five greater than the median $\chi^2$ for all stars on each line of sight.  We note that the limitations in our modeling of evolved stars are unlikely to be important here, because we eventually find all of the clouds in our catalog to reside nearer than 2.5~kpc, where dwarf stars are more prevalent than giant stars in the PS1 data.

We do not expect the simple dust screen model of Equation~\ref{eq:model} to be exact, and our individual star reddening estimates are occasionally catastrophically wrong due to problematic photometry and the presence of stars not well described by our models.  Accordingly we need to adopt a mechanism to reduce the influence of outliers on Equation~\ref{eq:palpha}.  We reduce the sensitivity of our algorithm to outliers by replacing each surface $p(E,D)$ with $p(E,D) + F$, giving any given star some chance of being an outlier.  We choose a value of $F$ so that stars in our model typically are found to be drawn from $p(E,D)$ with 75\% probability, and from the flat outlier distribution with 25\% probability.  The final distances we obtain are generally insensitive to $F$, except in a few cases where a catastrophically wrong solution is obtained when $F$ is too small.  For instance, for all but a few lines of sight, reducing $F$ by a factor of 100 changes the derived distances by less than 10\%.   However, our uncertainty estimates are very sensitive to $F$, as $F$ gives the fit freedom to ignore stars that would otherwise constrain the cloud's distance.  Reasonable values of $F$ can lead to statistical uncertainties in distance about half as large as those reported in this work; our values are conservative.  In general, systematic uncertainties dominate the error budget; see \textsection\ref{sec:sysunc}.

Some low Galactic latitude lines of sight have many stars within our $0.2\degree$ radius line of sight with $E(B-V) > 0.15$.  If needed, we limit the sample to 2000 stars along a single line of sight.  This speeds computation, which is otherwise dominated by lines of sight near Ophiuchus where large numbers of bulge stars are present.  In summary, we first select all stars in a 0.2\degree\ beam or all M-stars in 0.7\degree\ beam.  We then limit the selection to those stars with {\it Planck} $E(B-V) < 0.15$.  If more than 2000 stars pass these cuts, we select a random subsample of 2000 stars.  We then analyze each star to determine its reddening and distance probability distribution function, and exclude stars with large $\chi^2$.  The resulting stars are then used to determine the distance of each cloud.

\section{Results}
\label{sec:results}

We compile a catalog of distances to molecular clouds selected from two works: the CO maps of \citet{Dame:2001} and the catalog of \citet[MBM]{Magnani:1985}.  The MBM catalog provides specific sight lines through their clouds; we determine the distance to each cloud along those sight lines.  On the other hand, \citet{Dame:2001} provide maps of CO emission, rather than specific lines of sight.  That work labels many major cloud complexes, including Ophiuchus, Aquila South, Hercules, Lacerta, Pegasus, the Cepheus Flare, the Polaris Flare, Ursa Major, Camelopardis, Perseus, Taurus, $\lambda$ Orionis, Orion A and B, and Monoceros.  

We choose specific lines of sight that we deem to be representative of each cloud and suitable to our technique, with estimated $0.15 < E(B-V) < 3$ and spatially smooth $E(B-V)$ in the vicinity of the line of sight when possible.  These requirements allow a range of possible choices of lines of sight.  The locations we have chosen are often outside the parts of the cloud that have been subject to the most study, which tend to have large $E(B-V)$ unsuitable to our analysis.  In a few cases, we choose lines of sight through nearby clouds that our analysis later reveals to be unassociated with the main cloud of interest.  Users of this catalog should treat our categorization of lines of sight into regions like ``Orion'' with care; the sight line may be near, but outside of, the region traditionally associated with Orion.  We ignore the \citet{Dame:2001} clouds at low Galactic latitudes in the inner Galaxy, as robust modeling of these clouds would require accounting for the numerous dust clouds along each line of sight.  Clouds with $\delta < -30\degree$ are also excluded, as we have no PS1 photometry in that part of the sky. 

The MBM catalog is nicely matched to our PS1-based analysis.  That catalog is limited to high Galactic latitudes ($|b| > 20\degree$), and the sky observable from Texas, meaning that most MBM clouds have available PS1 photometry.  The MBM catalog provides fiducial coordinates for each cloud, so we fit $E(D)$ along these sight lines as given.  A small number of the MBM clouds reside at $\delta < -30\degree$ or where, because of bad weather, no PS1 photometry is yet available; these clouds are not included.  We additionally exclude sight lines where we find fewer than 10 nearby stars with Planck-estimated $E(B-V)$ more than $0.15$ magnitudes, as at this level imperfections in our stellar models and the PS1 photometry can masquerade as reddening signatures.  A few of the MBM clouds with $E(B-V) < 0.1$ appear to simply have no dust present, and are low-significance detections in the MBM catalog (e.g., MBM 10).

We use two different samples of stars in this analysis, depending on whether or not the cloud in question is close or far (\textsection\ref{sec:method}).  We use only nearby M-dwarfs and a large radius when the cloud is nearby, and we use all stars in a small radius when the cloud is distant.  We use the near technique for essentially all of the MBM clouds, and the far technique for the hand-selected clouds, with the following exceptions.  First, we use the near technique for sight lines through Ophiuchus, Aquila South, Hercules, and Taurus, as these are nearby clouds within about 200~pc.  Second, we use the far technique on MBM 46-48, which we find to be at a distance of about 480~pc.

\begin{figure*}[htb]
\dfplot{\detokenize{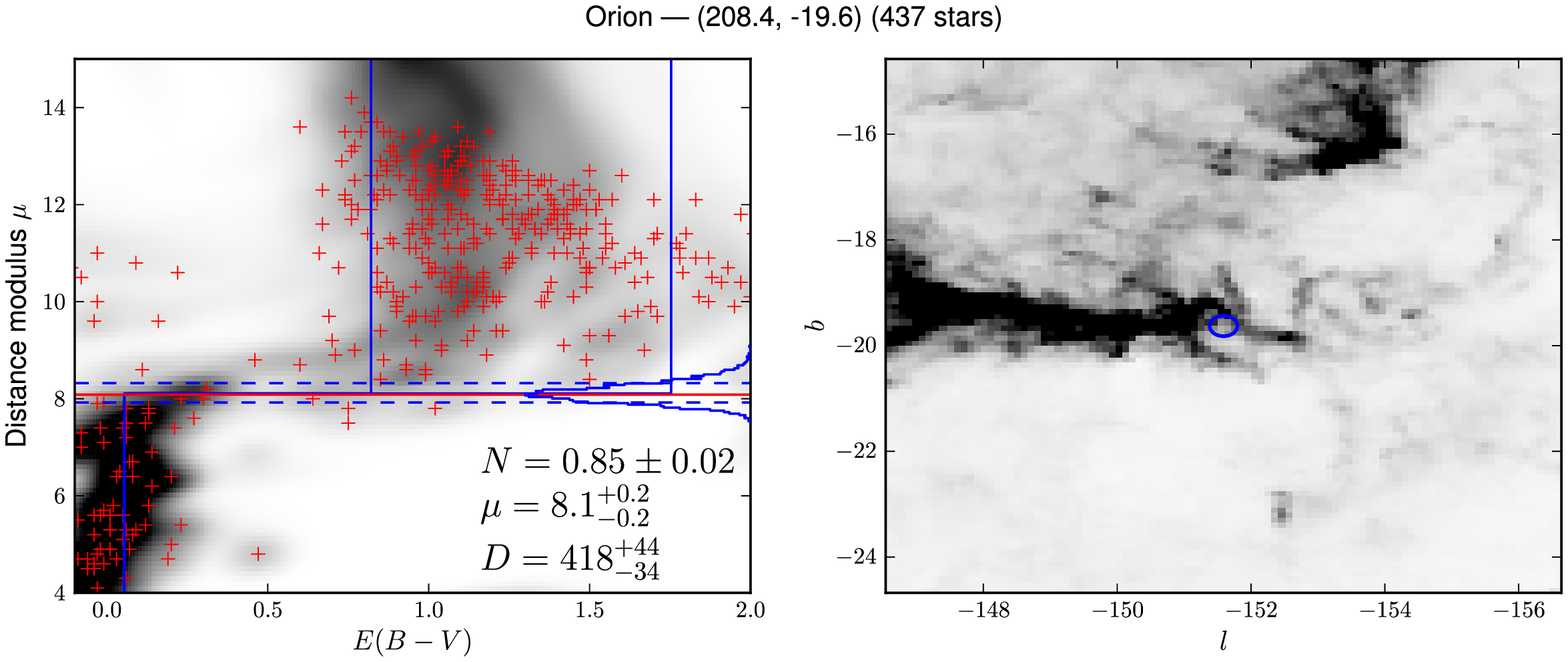}}
\dfplot{\detokenize{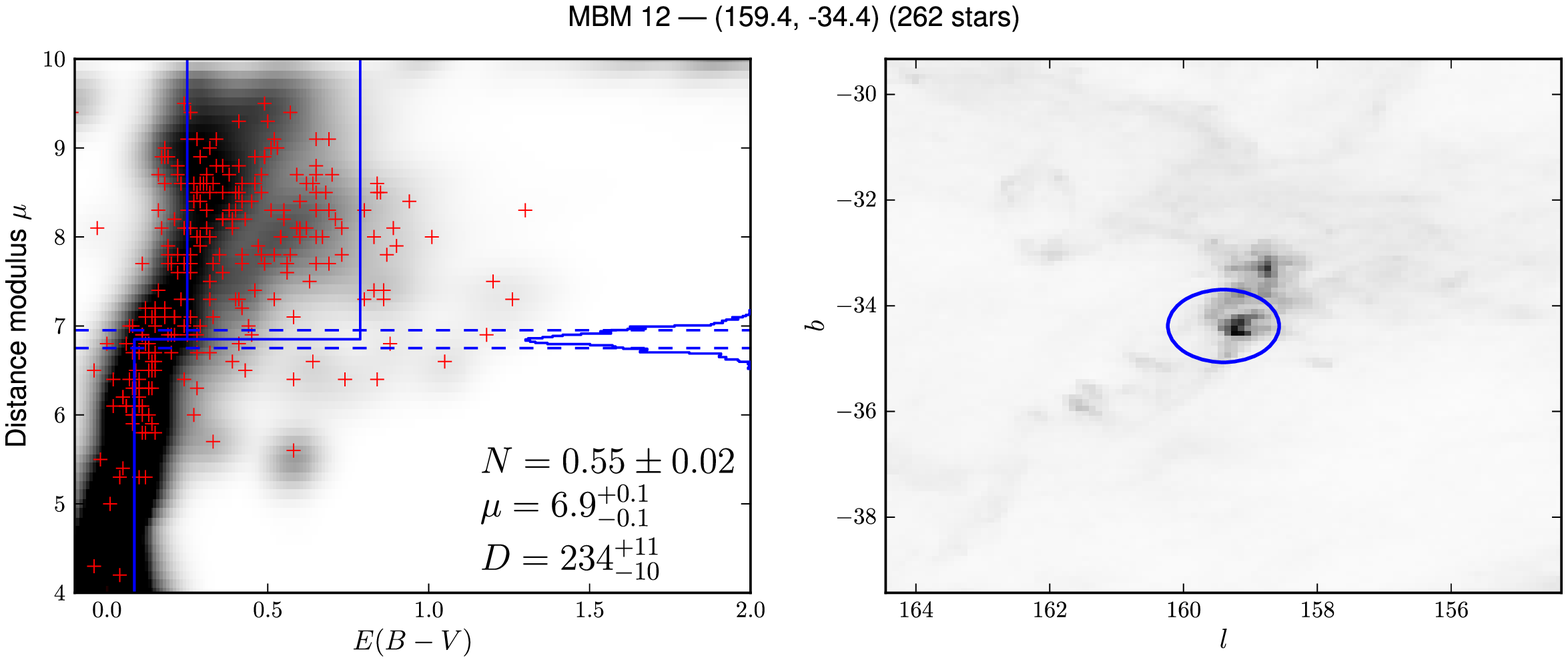}}
\dfplot{\detokenize{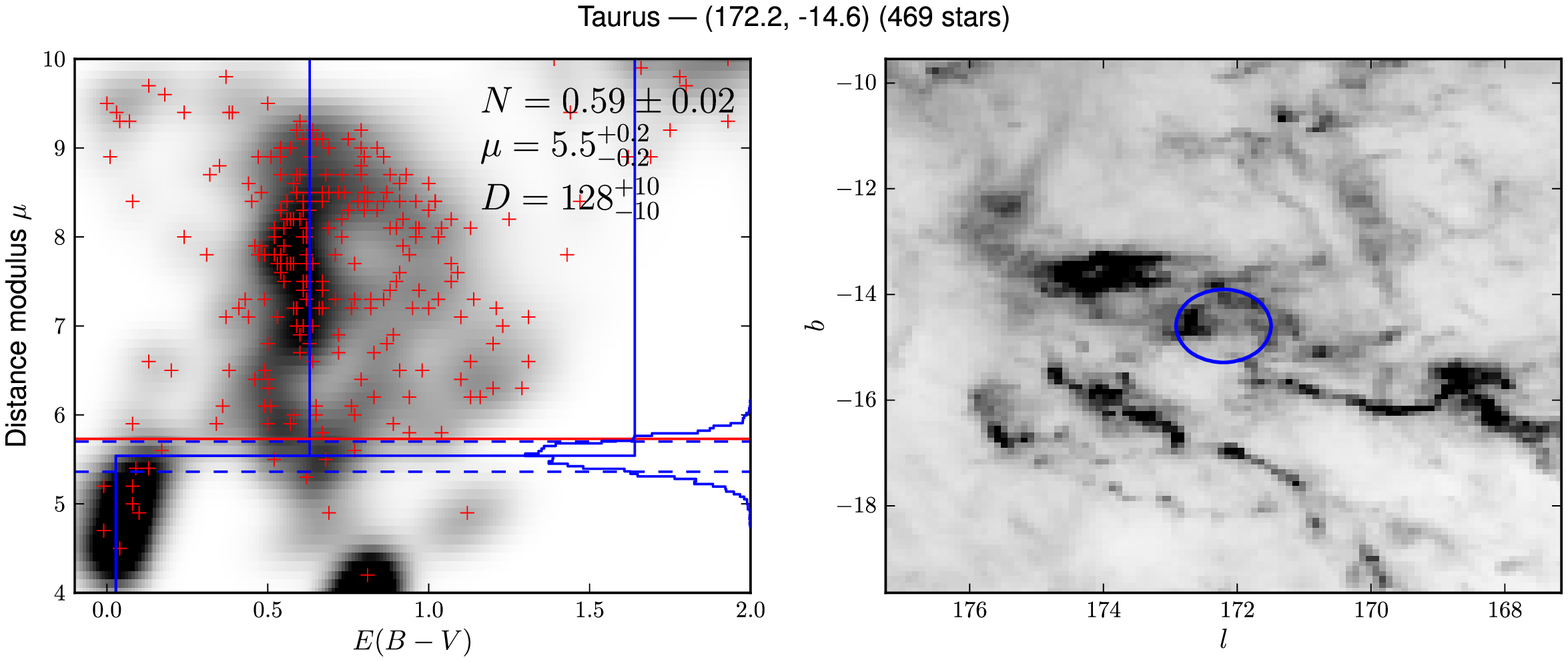}}
\figcaption{
\label{fig:distancefits} The distance to the Orion Nebula Cluster, MBM 12, and Taurus from \PS\ photometry.  The red crosses give the maximum likelihood reddenings and distances to the stars on each line of sight, and the underlying grayscales give the full probability distribution functions.  The blue histograms give the inferred probabilities of the possible distances to the cloud.  Blue dashed lines give the 16th and 84th percentiles, while the red lines gives literature distance estimates, when available.  The solid blue line gives the thin-dust screen fit.  See \textsection\ref{sec:results} for details.  The right hand panels shows the location of the line of sight in the context of the surrounding dust, as given by \citet{Planck:2011}.  The ellipse shows the approximate region of sky from which the stars in the analysis were drawn.  Our distance to the Orion Nebula Cluster and the Taurus Molecular Cloud are in good agreement with the \citet{Menten:2007} and \citet{Kenyon:1994} measurements, respectively.
}
\end{figure*}

We illustrate the distance determination in Figure \ref{fig:distancefits}, which shows the results of our analysis toward three lines of sight: toward the Orion Nebula Cluster (ONC) (top panels), MBM 12 (middle panels), and the Taurus molecular cloud (bottom panels).  The left panels show a visualization of the results of our fitting procedure.  We have computed $p(E,D)$ for each star and summed the results, showing them in the grayscale.  Nearby unreddened stars fall in the lower left of the panel, while distant reddened stars fall in the upper right.  We have additionally normalized the grayscale so that the same amount of weight falls into each distance modulus bin; there are many more stars around $\mu=10$ than with $\mu < 6$.  Red crosses show the maximum-likelihood locations for each of the stars on the line of sight.  The first panel shows the reddenings preferred by the stars at different distances.  The histogram in this panel shows the range of probable distances to the dust cloud on this line of sight, determined by the MCMC sampling of Equation \ref{eq:palpha}.  The blue line shows the reddening profile of Equation~\ref{eq:model}, using the median parameters from the MCMC sampling.  Beyond the cloud distance, the reddening profile splits into two branches, corresponding to the 16th and 84th percentile of the Planck dust map toward the stars in that region, times the median normalization the method obtains.  Shown in the upper right are the implied Planck dust map normalization factor $N$, the cloud distance modulus $\mu$, and the corresponding distance $D$ in pc.  We give the median values and the 16th and 84th percentiles through the uncertainties.  The right hand panel shows the Planck dust map in the region, with the approximate area the stars were drawn from given by the blue ellipse.  The label gives the Galactic latitude and longitude of the line of sight, as well as the number of stars used to constrain the distance.

The first row of panels in Figure~\ref{fig:distancefits} shows that along this sight line near the ONC, we find a cloud distance in good agreement with the work of \citet{Lombardi:2011}, shown on the figure by the red line.  This is also excellent agreement with the parallax distance of \citet{Menten:2007}.  We obtain $D = 418\pm39$~pc along this line of sight, compared with the parallax distance of $414\pm7$, though in general our distance estimates to Orion tend to be about $10\%$ higher than the parallax distance; see \textsection\ref{subsec:orion}.  The grayscale clearly shows that for $\mu < 8$, all stars are unreddened, while for $\mu > 8$, stars have reddenings predominantly near the expected reddening from the Planck map, though in detail we find that a reddening of $0.85$ times the Planck reddening provides a better fit.

The second row of panels shows the line of sight toward MBM 12.  We find that this cloud lies at about $D = 234$~pc.  At this close distance, we use the larger $0.7\degree$ beam to obtain a satisfactory number of foreground stars.  The relatively small reddening of the cloud makes the step at $D = 234$~pc less convincing; still, stars nearer than the step are generally unreddened, while those beyond it are not.

Finally, the bottom row of panels shows a line of sight through the Taurus molecular cloud, which is generally found to be at a distance of about 140~pc \citep{Kenyon:1994}.  Again, for this nearby cloud we use the ``near'' analysis to obtain enough foreground stars.  We find $D = 128\pm10$, in good agreement with the literature estimate.

For each sight line studied, we have made Figures analogous to Figure~\ref{fig:distancefits}, which are available in the online journal and at our web site\footnote{http://faun.rc.fas.harvard.edu/eschlafly/distances}.  However, as most clouds have approximately Gaussian distance probability distribution functions, we also tabulate the 16th, 50th, and 84th percentile of the probability distribution in Table~\ref{tab:clouddistances} for the major molecular clouds and in Table~\ref{tab:mbmdistances} for the MBM clouds.  We note that in a few cases (e.g., MBM 40), no obvious foreground stars are detected, and so our distance limits are determined by essentially the 16th and 84th percentiles of our flat prior on distance, out to the first observed reddened stars.

\begin{figure*}[htb]
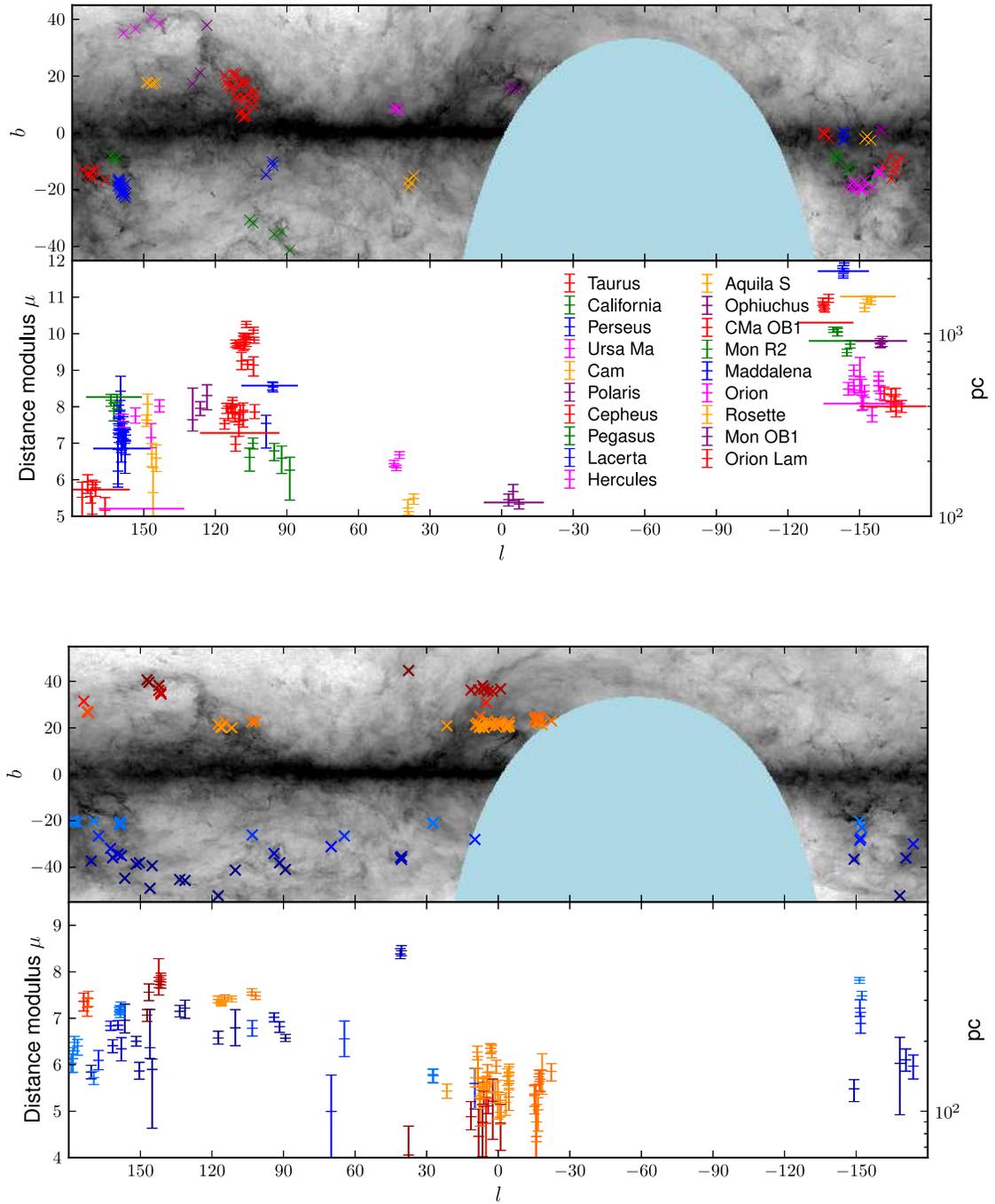

\dfplot{bigdistances_label}
\dfplot{mbmdistances}
\figcaption{
\label{fig:clouddistances} The distribution of lines of sight through major molecular clouds (top) and MBM molecular clouds (bottom), together with the inferred distances to the clouds and their uncertainties.  The top panels show the locations of the adopted lines of sight on a Planck-based map of dust, with sight lines colored by cloud complex (upper set of panels) or Galactic latitude (lower set).  The bottom panels show the distance modulus estimates to those clouds and their associated uncertainties, colored as in the top panels to aid association.  The right hand axes give the distances in physical units.  The uncertainties shown do not include the overall systematic uncertainties in the analysis.  See \textsection\ref{sec:results} for details.
}
\end{figure*}

We show the locations of the sight lines and the corresponding distances and statistical uncertainties in Figure~\ref{fig:clouddistances}.  The first set of panels shows the distribution of our hand-selected lines of sight through major molecular clouds, while the second set of panels shows the MBM catalog.  The upper panels show the sight lines overplotted on the Planck dust map, while the lower panels show the recovered distance moduli and their errors as a function of Galactic longitude.  In the first set of panels, points are colored by the cloud complex through which the sight lines were chosen, while in the second set, points are colored by their Galactic latitude.  This is intended to ease the association of points in the upper and lower panels.  The sight lines sample many of the most important dust clouds in the $\delta > -30\degree$ sky, yet they clearly cover the entire PanSTARRS-1 footprint only sparsely.  The lower panels of Figure~\ref{fig:clouddistances} shows the cloud distances that we measure, together with their uncertainties.  Literature distances to major molecular clouds are given by horizontal lines; references are in Table~\ref{tab:clouddistances}.  We measure cloud distances from $D \approx 100~\mathrm{pc}$ for Ophiuchus to $D \approx 2350~\mathrm{pc}$ for Maddalena's cloud.  

Our distances and statistical uncertainties are largely reliable.  The agreement between the literature distances and our determinations in Figure~\ref{fig:clouddistances} is good, and does not suggest a significant overall bias.  To give one example, our estimate of the distance to Orion Nebula Cluster (Figure~\ref{fig:distancefits}) is $418\pm43$~pc.  The \citet{Menten:2007} parallax distance to the Orion Nebula cloud is 414~pc, in close agreement.  However, we do find a systematically somewhat larger distance along most other sight lines through the Orion cloud; see \textsection\ref{subsec:orion} for details.  Our distance estimates for multiple sight lines through the same cloud are generally consistent, indicating that our uncertainites are reasonable.  For instance, the magenta points at $l \approx 150\degree$ corresponding to the Ursa Major molecular cloud are consistent within their uncertainties, and show similar distances to the neighboring purple and red points sampling Polaris and Cepheus.  Our distances to the clouds in Lacerta are mutually consistent, as are those in Pegasus.  We note that the uncertainties used in this figure do \emph{not} include the overall systematic uncertainties of 10\% we budget in our analysis (\textsection\ref{sec:sysunc}).

We have excluded from this analysis a few of the MBM clouds: those in Draco (MBM 41--44) and MBM 48 and 160.  These clouds are more distant than the typical MBM clouds, and we obtain badly inconsistent results when we fit them with the near and far techniques.  Draco may be problematic because of its status as an intermediate velocity cloud, which may lead dust content atypical of the Milky Way.  Alternatively, it may simply be problematic because of its relatively low reddening outside of two somewhat dense cores ($E(B-V) \approx 0.15$).  In the clouds MBM 48 and 160, the expected step in reddening at the distance to the cloud appears smeared out.  Accordingly we exclude these clouds from the catalog as well.  Several other MBM clouds are excluded because they lack any nearby stars with Planck-estimated $E(B-V) > 0.15$~mag or because they are outside of the \PS\ footprint.

\section{Systematic Uncertainties in Cloud Distance}
\label{sec:sysunc}

We often obtain statistical uncertainties in distance of less than 5\%.  Unfortunately, we adopt here an overall systematic uncertainty in distance of 10\%, and 15\% when using the ``near'' technique.  This systematic uncertainty is intended to account for a number limitations in our modeling: errors in our stellar models, inadequacy of our model for the dust, and variation in the dust reddening law.

\subsection{Stellar Models}
\label{subsec:stellarmodels}

Our distance estimates are ultimately tied to photometric distances to individual stars.  These are measured through the difference between their apparent magnitudes and their absolute magnitudes.  We only directly measure the apparent magnitudes; we obtain absolute magnitudes from the stars' colors through our library of intrinsic stellar colors, as described in \citet{Ivezic:2008}.  The relationship between colors and absolute magnitudes is also dependent on metallicity.  The \PS\ colors are largely insensitive to metallicity, and so we adopt a model for the distribution of stars' metallicities in space from \citet{Ivezic:2008}.  Any errors in these models will translate into systematic biases in our distance measurements.

The work of \citet{Yanny:2013} finds good agreement between the model of \citet{Ivezic:2008} and observations of the globular clusters NGC 2682 and NGC 2420, though they tweak the overall absolute luminosities by 0.06 magnitudes and additionally change the color dependence slightly.  The \PS\ data for these clusters shows similarly good agreement, though is not deep enough to reach late M stars.  The stellar models of \citet{Ivezic:2008} are designed to match old globular cluster populations and especially their main-sequence turn-off and higher-mass members, while in this work most of the stars considered are later types and part of a younger, more metal-rich disk population.  Our models also ignore binarity; unresolved identical binaries will have 40\% underestimated distances.  To address these limitations we adopt a more conservative 0.2~mag ($\sim 10\%$ distance) systematic uncertainty in the absolute magnitudes.  This systematic uncertainty applies to our overall distance scale, but not to the relative distances between different parts of the same region of a cloud.

\subsection{Photometric Calibration}
\label{subsec:cal}

A related systematic uncertainty is the photometric calibration of the survey.  All of the stars on each sight line are subject to the same photometric calibration errors, potentially leading to significant biases in derived distances and reddening.  The typical accuracy of the photometric calibration is expected to be better than 1\% in each band \citep{Schlafly:2012}.  We have experimented with introducing 2\% calibration erorrs in each of the bands in our sight lines through the California and Monoceros R2 clouds.  These simulated calibration errors induce changes in distance of always less than 5\%.  This suggests that photometric calibration errors are not dominant in this analysis.

\subsection{Dust Model}
\label{subsec:dust}

We make several assumptions about the dust in this analysis.  Spatially, we assume that the dust comes in a single thin screen with angular structure given by the dust map.  In its extinction, we assume that it is described by a single $R_V=3.1$ reddening law, and moreover that the reddening vector of \citet{Schlafly:2011} applies to all of the stars.  Both assumptions do not hold in detail.

Our dust screen model assumes that dust within the cloud of interest lies at a constant distance in a thin screen.  Given the current substantial uncertainties in the distances to many of these clouds, our simple model seems appropriate.  However, we have found substantial variations in distance to different components of the same cloud, clearly pointing to limitations in our approach.  Additionally, on several lines of sight our single-cloud model of reddening is violated, and at least two clouds lie on these sightlines.  This situation is not rare: in Cepheus, many sight lines contain reddening layers at both 300~pc and 900~pc; near Taurus, many sight lines have layers of reddening at both the distance of Taurus (140~pc) and Perseus (260~pc); and near Orion, we often see reddening associated with the Monoceros R2 reflection nebula (900~pc) and with Orion (400~pc).  This is not fatal because we choose sight lines where the dust column is dominated by the dust associated with only one of the clouds.  The 2D dust screens we adopt from Planck are also imperfect, though the use of other templates does not significantly change our derived distances.  We expect both of these limitations of our model to increase the uncertainties we derive in the distances to these clouds, rather than to bias our distance measurements.

Another limitation to our analysis is the assumption of a single $R_V = 3.1$ reddening law for all clouds.  However, dusty environments like the molecular clouds considered in this work have long been recognized to be associated with flatter $R_V \simeq 5$ reddening laws \citet{Cardelli:1989}.  For a cloud with reddening $E(B-V) = 1$, changing from $R_V = 3$ to $R_V = 5$ naively induces a change of 2~mag in distance modulus for stars behind the cloud, leading us to infer that the stars are about twice as far away as they actually are.  However, the work of \citet{Foster:2013} finds that when $E(B-V) \lesssim 0.7$, the standard $R_V=3.1$ reddening law is appropriate.  The flatter $R_V=5$ reddening law becomes appropriate at $A_V \simeq 10$ (i.e., $E(B-V) \simeq 2-3$, depending on $R_V$).  However, through our hand-selected lines of sight, in general $E(B-V) \lesssim 1$, and for the clouds of \citet{Magnani:1985} in general $E(B-V)$ is significantly lower than 1.  Moreover, because of the complex interaction between the reddening vector and the absolute magnitudes we infer for stars, in practice we find that variation in $R_V$ by one has only a minor effect on the distances we obtain.

We adopt the reddening vector of \citet{Schlafly:2011} for all stars in this analysis.  However, the reddening vector for a star depends on its type and, to a lesser extent, its reddening.  The reddening vector of \citet{Schlafly:2011} is appropriate only for unextinguished 7000~K F-stars.  For M-dwarfs, the reddening vector can be up to 15\% different in certain bands.  This suggests that our ``near'' analysis, which uses exclusively M-dwarfs, may be substantially biased.  However, because the reddening vector is nearly perpendicular to the stellar locus for these stars, we find changes in distance of less than 10\% when adopting a reddening vector more appropriate for M-dwarfs, and typically less than 5\%.

\subsection{Near vs. Far Analysis}
\label{subsec:nearfar}

We perform two very similar types of analysis in this work: a ``near'' analysis using color-selected M-dwarfs in a wide area, and a ``far'' analysis using all stars in a narrow area.  In general, the two types of analysis determine compatible distance estimates.  In the case of the northern component of the Cepheus Flare, however, our ``near'' technique determines a distance of approximately 300~pc, relative to the ``far'' technique which finds a distance of about 350~pc.  On the other hand, we do not see this effect as significantly in the Perseus Molecular Cloud or in Orion.  Nevertheless, we budget an additional 15\% systematic uncertainty in our ``near'' analysis.  A plausible explanation is that our model for the intrinsic colors of M-dwarfs is incomplete; our models neglect, for instance, the variation in M-dwarf color as a function of metallicity.  Our modeling also does not address the potential small bias in reddening stemming from the use only of objects that fall into our ``near'' color box (Equation~\ref{eq:brightred}), though varying the color box does not affect our final distances significantly.

\section{Discussion}
\label{sec:discussion}

The resulting catalog of distances is the largest homogeneous catalog of distances to molecular clouds available.  We span distances from 100~pc to 2400~pc, with typical uncertainties of only 10\%.

Some of the dust clouds in our analysis already have distance estimates from the literature.  In most cases, our distances are in good agreement with these preexisting estimates.  For example, our distances to Orion, Monoceros R2, and California agree with the estimates of \citet{Menten:2007}, \citet{Lombardi:2011} and \citet{Lada:2009}.  In the Polaris and Cepheus Flares there is substantial discrepancy between various literature distance estimates; we find that both of these complexes share a distance of about 370~pc.  We further place Ursa Major at the same distance, substantially revising the too-nearby distance of 110~pc advocated by \citet{Penprase:1993}.  For many clouds, no estimates are available, and the measurements we make are the first.  However, given the absence of large catalogs of cloud distances in the literature, it can be challenging to find estimates to particular clouds.  We have made a serious effort to find literature distances only to a selection of major molecular clouds, and not to each the MBM clouds---often which are themselves members of the same major molecular clouds.

In principle, our technique is sensitive to clouds out to distances as large as 5~kpc, limited by the distance out to which PS1 provides good 5-color photometry of significantly reddened main-sequence stars.  The most distant cloud in our sample is Maddalena's cloud, at a distance of about 2350~pc.  However, at high Galactic latitudes the majority of clouds are nearby, consistent with expectations of a $\sim 100$~pc scale height of the gas in the Galaxy \citep{Kalberla:2009}.

We remark below on our results for a few specific major dust clouds.  Throughout we give a 10\% systematic distance uncertainty floor in our reported distances to clouds studied with the ``far'' technique.  When using the ``near'' technique, we give use a 15\% uncertainty floor.  The specific values reported below are syntheses of the results for individual lines of sight given in Table~\ref{tab:clouddistances} and Table~\ref{tab:mbmdistances} and are not strictly weighted averages of those results.

\subsection{Ursa Major, Polaris, and the Cepheus Flare}
\label{subsec:ursamajor}
\label{subsec:polarisflare}
\label{subsec:cepheusflare}

We find approximately common distances to these three major high Galactic latitude clouds, with distances of about 370~pc.  This conclusion is not surprising given their apparent relation in projected two-dimensional maps, but it is nevertheless remarkable given the $\approx 60\degree$ angular extent of the structure.  The eastern tip of Ursa Major has a distance within about 50~pc of the southwestern edge of the near Cepheus Flare, despite being physically separated by 300~pc at our adopted distance to the cloud.

The distance to the Ursa Major molecular cloud individually has not been extensively studied.  The work of \citet{Penprase:1993} places the cloud MBM 30 at a distance of $110 \pm 10$~pc, which \citet{Pound:1997} adopt as the distance to the Ursa Major molecular cloud.  We find the much larger distance of about $D = 350 \pm 35$~pc, in common with the Polaris and Cepheus Flares.

The distance to the Polaris Flare has been the focus of more study.  The first estimates of its distance are from \citet{Keenan:1941}, who place the cloud between 200 and 400~pc.  \citet{Heithausen:1990} place the cloud at 240 pc, largely to place it at the same distance as the Cepheus Flare, which \citet{Grenier:1989} place at 250--300 pc.  \citet{Zagury:1999} assumes that Polaris is behind the Polaris Flare, placing the cloud at a significantly closer distance of about 100 pc.  \citet{Brunt:2003} puts the cloud at a distance of $205 \pm 62$ pc.  Our own distances estimate is $D = 380 \pm 40$~pc, placing it at about the same distance as our estimates for the Cepheus Flare.  We note that this implies that Polaris is not reddened by the Polaris Flare.  We note that the current processing of the \PS\ data excludes data north of $\delta = 80\degree$, excluding the bulk of the Polaris Flare; the lines we study in its vicinity are through the outskirts of the Flare.

The Cepheus Flare is a complicated structure reviewed in detail in \citet{Kun:2008}.  That review makes clear that the Cepheus Flare contains a number of different components, as identified by velocity information from the molecular gas.  Our analysis separates the cloud into two components at different distances, without knowledge of the velocity information.  These two components are illustrated by the distinct groups of red points around $l = 110\degree$ in Figure~\ref{fig:clouddistances}.  The northern component, with $95 < l < 115$ and $b > 14.5$, is the more nearby, with a distance of $D = 360 \pm 35$.  This distance is somewhat larger than the literature distances to the Flare.  For example, \citet{Kun:1993} give $300^{+50}_{-10}$ pc to the associated L1241 and \citet{Zdanavicius:2011} measure $286 \pm 20$ pc for the distance to a portion of the cloud at $(l, b) = (102.5, 15.5)$.

On the other hand, our analysis identifies the southern component of the Flare, with $105 < l < 115$ and $11.5 < b < 14.5$, as much more distant, with $D = 900 \pm 90$~pc.  The distance to this component of the cloud was described by \citet{Kun:2008} as controversial; our technique clearly identifies the distance to this cloud.  \citet{Kiss:2006} present maps of the distances to clouds in Cepheus that likewise clearly separate the northern and southern components of the Flare.  That work adopts a southern component distance of only about 400~pc, incompatible with our value.  Our separation into two clouds at distances of 900~pc and 360~pc is close to the separation of \citet{Grenier:1989}, who identify far and near components at 800 and 250~pc.

We note that because of the presence of two clouds in this region, our technique is not completely suitable.  Some of our adopted lines of sight show significant extinction at the distance to each of the two clouds.  In general, along these lines the most nearby unreddened stars are deemed outliers, and the more distant cloud is deemed the single cloud along the line of sight.  In a few cases, the algorithm adopts a spurious distance between the two clouds, as seen in Figure~\ref{fig:clouddistances}.

\subsection{Camelopardis}
\label{subsec:camelopardis}

Three sight lines through clouds in Camelopardis are assigned distances of about $200 \pm 30$~pc.  Two lines of sight through a plausibly related adjacent cloud are determined to have distances $D = 350 \pm 35$~pc, considerably more distant.  We are aware of no literature estimates of the distance to these clouds.  The more distant cloud has distance similar to the nearby Ursa Major-Polaris-Cepheus molecular cloud complex.

\subsection{Taurus}
\label{subsec:taurus}

The Taurus molecular cloud is among the most nearby giant molecular clouds.  Accordingly, for lines of sight through this cloud we use our ``near'' technique, with larger $0.7\degree$ beams and only M-stars.  The result is a distance estimate of $135\pm20$~pc, compatible with the literature estimate of about 140~pc (e.g., \citet{Kenyon:1994}).

\subsection{Perseus}
\label{subsec:perseus}

The Perseus Molecular Cloud is the most actively star-forming cloud within 300~pc from the sun \citep{Bally:2008}. Literature estimates of the cloud distance range from 350~pc \citep{Herbig:1983} to parallax estimates of about 230~pc from \citet{Hirota:2008} and \citet{Hirota:2011}.  A significant velocity gradient exists across the cloud, suggesting that different parts of the cloud may lie at different distances \citep{Bally:2008}.  \citet{Cernis:1990} and \citet{Cernis:1993} find a distance gradient across the cloud, obtaining a distance of 220~pc to the western component of the cloud and 260--340~pc to the eastern component.  We find a similar result: lines of sight through the western portion of the cloud are consistent with $260 \pm 26$~pc, while lines of sight through the eastern portion are consistent with $D = 315 \pm 32$~pc.

\subsection{California}
\label{subsec:california}

We study three lines of sight through the outskirts of the California molecular cloud.  The estimates are compatible and find $D = 410 \pm 41$~pc.  The work of \citet{Lada:2009} finds $D = 450 \pm 23$~pc, in good agreement.  We accordingly agree with the assessment of \citet{Lada:2009} that the California molecular cloud is of comparable mass to the Orion giant molecular cloud.

\subsection{Pegasus}
\label{subsec:pegasus}

We place several molecular clouds in Pegasus at a distance of $D = 230 \pm 23$~pc, for which we are aware of no literature distance estimate.

\subsection{Lacerta}
\label{subsec:lacerta}

We place the Lacerta molecular cloud at a distance of $D = 510 \pm 51$~pc.  This is in mild tension with the distance of $360 \pm 65$ adopted to the associated cloud LBN 437 by \citet{Soam:2013}, but is compatible with the estimated distance of $520\pm20$~pc distance estimated to the OB association Lacerta OB1 by \citet{Kaltcheva:2009}.  We also placed one line of sight through the relatively diffuse dust trailing Lacerta; this cloud turns out to be much more nearby ($D = 300 \pm 60$~pc), unassociated with the main cloud at $95\degree < l < 97\degree$, $-15\degree < b < -9\degree$.

\subsection{Hercules}
\label{subsec:hercules}

This dust cloud has been mapped in CO by \citet{Dame:2001}, but we are unaware of a previous estimate of its distance.  Taking three lines of sight through the cloud, we obtain a consistent distance of $D = 200\pm30$~pc using our ``near'' technique.

\subsection{Aquila South}
\label{subsec:aquilasouth}

Like Hercules, this dust cloud in included in the maps of \citet{Dame:2001}, but we cannot find an estimate of its distance in the literature.  We find $D = 110\pm15$~pc, making this cloud among the most nearby molecular clouds.  Due to the cloud's proximity, we use the ``near'' technique for this cloud.

\subsection{Ophiuchus}
\label{subsec:ophiuchus}

We obtain a distance of $D = 125 \pm 18$ to Ophiuchus using the ``near'' technique.  This well-studied cloud is believed to reside at a distance of 119~pc \citep{Lombardi:2008}, consistent with our measurement.  We note that on many sight lines through this complex, there are clear signals of multiple layers of extinction in the region, with one at approximately 120~pc and the other at approximately 180~pc.  

\subsection{Orion}
\label{subsec:orion}

The Orion Molecular Cloud may be the most extensively studied molecular cloud in the Galaxy.  Literature distance estimates include parallaxes to stars in the Orion Nebula Cluster, which find $D = 414 \pm 7$~pc \citep{Menten:2007} and $389 \pm 23$~pc \citep{Sandstrom:2007}.  The star-count based estimates of \citet{Lombardi:2011} find a similar value of $371 \pm 10$~pc.

We study several lines of sight through the Orion A complex and the $\lambda$ Orionis molecular ring, in order to get a good handle on the distance to this important cloud.  Because of the large extinction through the center of the cloud, we choose sight lines through the outskirts of the cloud---a somewhat risky procedure owing to the presence of the more distant Monoceros R2 complex in the vicinity.  Choosing two lines of sight near the ONC, we find $D = 420\pm42$~pc, in good agreement.  We find the same distance toward clouds in the $\lambda$ Orionis ring.

However, in general our lines of sight through Orion give larger estimates for Orion's distance, more consistent with $D = 490\pm50$~pc on five independent lines.  Likewise we favor a larger distance to the Scissors, with $D = 520\pm52$~pc, suggesting a complicated three-dimensional structure in Orion.  This suggests, for instance, that the eastern edge of Orion is 70~pc farther from us than the ONC, compared to its approximately 30~pc extent in projection.

\subsection{Monoceros R2}
\label{subsec:monr2}

This cloud was determined by \citet{Lombardi:2011} to reside at a distance of $905 \pm 37$~pc, which refined an older estimate of $830 \pm 50$~pc \citep{Herbst:1976}.  We find that the clouds near the core of the complex have a distance of $D=830\pm83$~pc, while the ``Crossbones'' toward the northeastern edge have a greater distance of $D = 1040\pm104$~pc.  The physical separation in the plane of the sky between these two parts of Monoceros is about 100~pc, significantly smaller than the $\sim 200$~pc difference we find in distance along the line of sight.

\subsection{MBM Clouds}
The review of \citet{McGehee:2008} summarizes findings about the properties of a selection of interesting high Galactic latitude MBM clouds.  We can comment on each of these.

The cloud MBM 7 is a translucent cloud estimated by \citet{Magnani:1986} to have distance $75 < D < 175$~pc.  We find $D = 148\pm20$~pc, consistent with but improving that estimate.

The cloud MBM 12 was once believed to be the most nearby molecular cloud, provoking substantial interest.  More recent distance estimates range from 275~pc \citep{Luhman:2001} to 360~pc \citep{Andersson:2002}.  Our method determines $D = 234\pm35$~pc, in agreement with the distance of \citep{Luhman:2001} but in tension with the result of \citep{Andersson:2002}.  In either case, we agree with the recent literature in that this cloud is not the nearest molecular cloud.  

Among the clouds considered in this work, instead the cloud MBM 40 seems to be the closest.  We detect no stars in the foreground of that cloud, though that is probably largely due to the small area of sky nearby that has $E(B-V) > 0.15$~mag.  The closest stars in the direction of MBM 40 require that the cloud lie at $D < 85$~pc (84th percentile).  

The cloud MBM 16 was estimated by \citep{Hobbs:1988} to lie at a distance of $60 < D < 95$~pc.  We find $D = 147\pm22$~pc, in mild tension with that result.

The work of \citet{Hearty:2000} finds a distance of about $110 < D < 170$~pc to the cloud MBM 20.  Our method determines $D = 124\pm19$~pc, consistent with that measurement.  

The MBM catalog contains many clouds in the vicinity of Ophiuchus.  The higher latitude clouds tend to be more nearby than the more distant ones, consistent with the picture of the tilted local chimney advocated by \citet{Lallement:2014}.  Likewise, as suggested by \citet{Lallement:2014}, there seem to be at least two nearby layers of extinction toward the Galactic center: one at approximately 110~pc and a second at approximately 180~pc.  These two layers are clearly evident in the our analysis of the MBM 135 and MBM 138 clouds, though it is relatively generic in the sight lines with $|l| < 30\degree$ and $|b| < 25\degree$.  

\section{Conclusion}
\label{sec:conclusion}

We present a catalog of distances to molecular clouds.  We obtain secure distance estimates to most of the high-latitude MBM molecular clouds, with accuracy typically limited by systematics to 15\%.  We further obtain distances to a number of well-studied molecular clouds at high latitudes or in the outer Galaxy.  We obtain secure distances to Lacerta, Pegasus, the Cepheus Flare, the Polaris Flare, Ursa Major, Camelopardis, Perseus, Taurus, $\lambda$ Orionis, Orion A, and Monoceros R2, in general confirming but refining the accepted distances to these clouds.  We highlight the complexity of the Cepheus Flare region, separating the molecular gas there into nearby (360~pc) and distant (900~pc) parts.  We correct the literature distance to Ursa Major and clarify the distance to the Polaris Flare, placing them at the distance of the nearby component of the Cepheus Flare.  We make the first distance estimates of which we are aware for a number of clouds, including clouds in Camelopardis and Pegasus, as well as a number of the MBM clouds.  Our distance estimates reach statistical uncertainties better than 5\% in 0.2\degree\ radius lines of sight, though we caution that we only expect absolute accuracies of 10\% owing to systematic uncertainties in our technique.  

This work has only scraped the surface of what is possible with the combination of this technique and the PS1 photometry.  We have focused on a sampling of sight lines through well-studied molecular clouds, but in principle the entire $\delta > -30\degree$ sky is amenable to this analysis.  The 5\% relative distance accuracy suggests that we can make 3D maps of major molecular clouds.  At the distance of Orion, we would expect to be able to obtain 20~pc distance resolution, compared with the $\sim 80$~pc projected angular size of the Orion A and B complex.  This initial work has aimed only to get accurate overall distances to major molecular clouds, but already hints of distance gradients across clouds like Orion and Perseus are present.

Moreover, the data used in this study come from the PS1 single-epoch data.  The PS1 Science Consortium is rapidly improving deeper data coming from stacks of the $\sim 7$ PS1 exposures of each part of the sky in each filter.  This stacking process increases the limiting magnitude of the survey by about a magnitude, allowing access to stars 50\% further away than we currently consider.  Additionally, PS1 parallax and proper motion studies are beginning to bear fruit, opening the possibility of incorporating astrometric information into our distance estimates.  Such an effort would dramatically improve our distance and reddening measurements to individual stars and serve as a useful pathfinder to the Gaia mission.

We additionally look forward to eventually including other sources of data in our analysis.  Our technique is naturally extendable to accept other sources of photometry.  Of particular interest is infrared photometry, which, as demonstrated by \citet{Lombardi:2011}, can strongly constrain the distances to molecular clouds.  Infrared data would allow us to probe higher column density clouds and improve our ability to discriminate variation in stellar temperature from variation in reddening.  Existing data from 2MASS would already be helpful; upcoming infrared surveys like the UKIRT Hemisphere Survey and the VISTA Hemisphere Survey, which are better matched to the depth of PS1, will provide still better leverage.

ES acknowledges support from the DFG grant SFB 881 (A3).  GMG and DPF are partially supported by NSF grant AST-1312891.  N.F.M. gratefully acknowledges the CNRS for support through PICS project PICS06183.  This research has made use of the SIMBAD database, operated at CDS, Strasbourg, France.

The Pan-STARRS1 Surveys (PS1) have been made possible through contributions of the Institute for Astronomy, the University of Hawaii, the Pan-STARRS Project Office, the Max-Planck Society and its participating institutes, the Max Planck Institute for Astronomy, Heidelberg and the Max Planck Institute for Extraterrestrial Physics, Garching, The Johns Hopkins University, Durham University, the University of Edinburgh, Queen's University Belfast, the Harvard-Smithsonian Center for Astrophysics, the Las Cumbres Observatory Global Telescope Network Incorporated, the National Central University of Taiwan, the Space Telescope Science Institute, the National Aeronautics and Space Administration under Grant No. NNX08AR22G issued through the Planetary Science Division of the NASA Science Mission Directorate, the National Science Foundation under Grant No. AST-1238877, the University of Maryland, and Eotvos Lorand University (ELTE).

\bibliography{2dmap}

\LongTables
\tabletypesize{\scriptsize}
\begin{deluxetable*}{ccccccc|ccccccc}
\tablecaption{Distances to Molecular Clouds
\label{tab:clouddistances}
}
\tablehead{
\colhead{Cloud} & \colhead{$l$ (\degree)} & \colhead{$b$ (\degree)} & \colhead{$E$} & \colhead{$N$} & \colhead{$D$ (pc)} & $D_\mathrm{lit}$ (pc) & \colhead{Cloud} & \colhead{$l$ (\degree)} & \colhead{$b$ (\degree)} & \colhead{$E$} & \colhead{$N$} & \colhead{$D$ (pc)} & $D_\mathrm{lit}$ (pc)
}
\startdata

       Aquila S &   37.8 &  -17.5 &  0.40 & $ 0.69^{+ 0.05}_{- 0.04}$ & $   76^{+   16}_{-   21}$ &                            &       Ophiuchus &  355.2 &   16.0 &  0.97 & $ 0.50^{+ 0.02}_{- 0.02}$ & $  136^{+   12}_{-   14}$ &   119\tablenotemark{    8} \\
       Aquila S &   38.9 &  -19.1 &  0.29 & $ 0.78^{+ 0.06}_{- 0.07}$ & $   89^{+   16}_{-   22}$ &                            &       Ophiuchus &  352.7 &   15.4 &  1.04 & $ 0.50^{+ 0.02}_{- 0.02}$ & $  116^{+    7}_{-    7}$ &   119\tablenotemark{    8} \\
       Aquila S &   36.8 &  -15.1 &  0.41 & $ 0.61^{+ 0.04}_{- 0.03}$ & $  125^{+    7}_{-    8}$ &                            &       Ophiuchus &  357.1 &   15.7 &  0.73 & $ 0.60^{+ 0.02}_{- 0.02}$ & $  123^{+    9}_{-    9}$ &   119\tablenotemark{    8} \\
       Aquila S &   39.3 &  -16.8 &  0.52 & $ 0.68^{+ 0.03}_{- 0.03}$ & $  111^{+   12}_{-    9}$ &                            &           Orion &  208.4 &  -19.6 &  1.22 & $ 0.85^{+ 0.02}_{- 0.02}$ & $  418^{+   43}_{-   34}$ &   414\tablenotemark{    9} \\
        CMa OB1 &  224.5 &   -0.2 &  1.63 & $ 0.50^{+ 0.01}_{- 0.01}$ & $ 1369^{+   64}_{-   56}$ &  1150\tablenotemark{    1} &           Orion &  202.0 &  -13.3 &  0.90 & $ 0.76^{+ 0.02}_{- 0.02}$ & $  519^{+   35}_{-   34}$ &   414\tablenotemark{    9} \\
        CMa OB1 &  222.9 &   -1.9 &  1.76 & $ 0.57^{+ 0.01}_{- 0.01}$ & $ 1561^{+   79}_{-   77}$ &  1150\tablenotemark{    1} &           Orion &  212.4 &  -17.3 &  0.65 & $ 0.68^{+ 0.03}_{- 0.03}$ & $  629^{+   43}_{-   43}$ &   414\tablenotemark{    9} \\
        CMa OB1 &  225.0 &   -0.2 &  2.02 & $ 0.52^{+ 0.01}_{- 0.01}$ & $ 1398^{+   63}_{-   59}$ &  1150\tablenotemark{    1} &           Orion &  201.3 &  -13.8 &  0.56 & $ 0.73^{+ 0.03}_{- 0.03}$ & $  470^{+   49}_{-   33}$ &   414\tablenotemark{    9} \\
        CMa OB1 &  225.4 &    0.3 &  1.21 & $ 0.56^{+ 0.01}_{- 0.01}$ & $ 1494^{+   72}_{-   66}$ &  1150\tablenotemark{    1} &           Orion &  209.8 &  -19.5 &  3.13 & $ 0.55^{+ 0.06}_{- 0.02}$ & $  580^{+  161}_{-  107}$ &   414\tablenotemark{    9} \\
     California &  163.8 &   -7.9 &  1.09 & $ 0.68^{+ 0.02}_{- 0.01}$ & $  431^{+   33}_{-   31}$ &   450\tablenotemark{    2} &           Orion &  214.7 &  -19.0 &  1.21 & $ 0.82^{+ 0.02}_{- 0.02}$ & $  497^{+   42}_{-   36}$ &   414\tablenotemark{    9} \\
     California &  161.2 &   -9.0 &  1.72 & $ 0.71^{+ 0.01}_{- 0.01}$ & $  421^{+   43}_{-   34}$ &   450\tablenotemark{    2} &           Orion &  207.9 &  -16.8 &  0.90 & $ 0.86^{+ 0.02}_{- 0.02}$ & $  484^{+   37}_{-   35}$ &   414\tablenotemark{    9} \\
     California &  162.5 &   -9.5 &  0.72 & $ 0.88^{+ 0.04}_{- 0.06}$ & $  377^{+   39}_{-   44}$ &   450\tablenotemark{    2} &           Orion &  212.4 &  -19.9 &  1.41 & $ 0.75^{+ 0.01}_{- 0.01}$ & $  517^{+   44}_{-   38}$ &   414\tablenotemark{    9} \\
            Cam &  146.1 &   17.7 &  0.67 & $ 0.60^{+ 0.03}_{- 0.04}$ & $  134^{+   50}_{-   36}$ &                            &           Orion &  212.2 &  -18.6 &  1.02 & $ 0.79^{+ 0.02}_{- 0.02}$ & $  490^{+   27}_{-   27}$ &   414\tablenotemark{    9} \\
            Cam &  144.8 &   17.8 &  0.43 & $ 0.71^{+ 0.06}_{- 0.08}$ & $  208^{+   37}_{-   32}$ &                            &           Orion &  209.1 &  -19.9 &  1.60 & $ 0.95^{+ 0.02}_{- 0.02}$ & $  478^{+   84}_{-   59}$ &   414\tablenotemark{    9} \\
            Cam &  148.4 &   17.7 &  0.33 & $ 0.73^{+ 0.07}_{- 0.06}$ & $  410^{+   56}_{-   86}$ &                            &           Orion &  209.0 &  -20.1 &  1.10 & $ 0.94^{+ 0.02}_{- 0.02}$ & $  416^{+   42}_{-   36}$ &   414\tablenotemark{    9} \\
            Cam &  148.8 &   17.8 &  0.30 & $ 0.77^{+ 0.07}_{- 0.07}$ & $  336^{+   22}_{-   25}$ &                            &           Orion &  202.0 &  -14.0 &  0.64 & $ 0.69^{+ 0.03}_{- 0.03}$ & $  585^{+   32}_{-   36}$ &   414\tablenotemark{    9} \\
            Cam &  146.6 &   17.2 &  0.43 & $ 0.71^{+ 0.02}_{- 0.04}$ & $  218^{+   31}_{-   32}$ &                            &           Orion &  204.7 &  -19.2 &  0.23 & $ 0.77^{+ 0.02}_{- 0.03}$ & $  356^{+   37}_{-   27}$ &   414\tablenotemark{    9} \\
        Cepheus &  106.4 &   17.7 &  0.79 & $ 0.73^{+ 0.02}_{- 0.01}$ & $  678^{+   39}_{-   41}$ &   286\tablenotemark{    3} &       Orion Lam &  196.7 &  -16.1 &  0.87 & $ 0.73^{+ 0.01}_{- 0.01}$ & $  461^{+   37}_{-   35}$ &   400\tablenotemark{   10} \\
        Cepheus &  110.7 &   12.6 &  0.94 & $ 0.67^{+ 0.01}_{- 0.01}$ & $  859^{+   33}_{-   30}$ &   286\tablenotemark{    3} &       Orion Lam &  194.7 &  -10.1 &  0.57 & $ 0.87^{+ 0.01}_{- 0.01}$ & $  378^{+   31}_{-   28}$ &   400\tablenotemark{   10} \\
        Cepheus &  108.3 &   12.4 &  0.76 & $ 0.77^{+ 0.01}_{- 0.01}$ & $  971^{+   37}_{-   39}$ &   286\tablenotemark{    3} &       Orion Lam &  195.5 &  -13.7 &  0.40 & $ 0.66^{+ 0.05}_{- 0.04}$ & $  425^{+   27}_{-   25}$ &   400\tablenotemark{   10} \\
        Cepheus &  105.9 &   13.8 &  0.76 & $ 0.68^{+ 0.01}_{- 0.01}$ & $  961^{+   29}_{-   34}$ &   286\tablenotemark{    3} &       Orion Lam &  194.8 &  -12.1 &  0.39 & $ 0.83^{+ 0.05}_{- 0.05}$ & $  463^{+   38}_{-   42}$ &   400\tablenotemark{   10} \\
        Cepheus &  107.0 &    9.4 &  0.82 & $ 0.60^{+ 0.02}_{- 0.02}$ & $ 1124^{+   43}_{-   43}$ &   286\tablenotemark{    3} &       Orion Lam &  196.9 &   -8.2 &  0.71 & $ 0.82^{+ 0.01}_{- 0.01}$ & $  397^{+   26}_{-   23}$ &   400\tablenotemark{   10} \\
        Cepheus &  107.0 &    6.0 &  1.52 & $ 0.52^{+ 0.01}_{- 0.01}$ & $  897^{+   50}_{-   46}$ &   286\tablenotemark{    3} &       Orion Lam &  199.6 &  -11.9 &  1.27 & $ 0.88^{+ 0.01}_{- 0.01}$ & $  468^{+   41}_{-   38}$ &   400\tablenotemark{   10} \\
        Cepheus &  103.7 &   11.4 &  0.85 & $ 0.61^{+ 0.01}_{- 0.01}$ & $  923^{+   33}_{-   35}$ &   286\tablenotemark{    3} &       Orion Lam &  192.3 &   -8.9 &  0.77 & $ 0.88^{+ 0.02}_{- 0.02}$ & $  400^{+   30}_{-   29}$ &   400\tablenotemark{   10} \\
        Cepheus &  108.4 &   18.6 &  0.85 & $ 0.81^{+ 0.03}_{- 0.03}$ & $  338^{+   42}_{-   30}$ &   286\tablenotemark{    3} &         Pegasus &  104.2 &  -31.7 &  0.27 & $ 0.84^{+ 0.05}_{- 0.06}$ & $  250^{+   17}_{-   17}$ &                            \\
        Cepheus &  109.6 &    6.8 &  2.28 & $ 0.47^{+ 0.01}_{- 0.01}$ & $  865^{+   39}_{-   35}$ &   286\tablenotemark{    3} &         Pegasus &  105.6 &  -30.6 &  0.28 & $ 0.82^{+ 0.06}_{- 0.08}$ & $  210^{+   25}_{-   33}$ &                            \\
        Cepheus &  108.2 &    5.5 &  1.45 & $ 0.57^{+ 0.01}_{- 0.01}$ & $  853^{+   48}_{-   44}$ &   286\tablenotemark{    3} &         Pegasus &   92.2 &  -34.7 &  0.46 & $ 0.63^{+ 0.06}_{- 0.07}$ & $  207^{+   34}_{-   36}$ &                            \\
        Cepheus &  107.7 &    5.9 &  1.39 & $ 0.53^{+ 0.01}_{- 0.01}$ & $  874^{+   58}_{-   52}$ &   286\tablenotemark{    3} &         Pegasus &   95.3 &  -35.7 &  0.35 & $ 0.79^{+ 0.06}_{- 0.07}$ & $  228^{+   22}_{-   26}$ &                            \\
        Cepheus &  104.0 &    9.4 &  1.10 & $ 0.63^{+ 0.02}_{- 0.01}$ & $ 1045^{+   40}_{-   38}$ &   286\tablenotemark{    3} &         Pegasus &   88.8 &  -41.3 &  0.53 & $ 0.66^{+ 0.06}_{- 0.07}$ & $  178^{+   31}_{-   56}$ &                            \\
        Cepheus &  109.6 &   16.9 &  0.79 & $ 0.92^{+ 0.03}_{- 0.04}$ & $  369^{+   36}_{-   36}$ &   286\tablenotemark{    3} &         Perseus &  160.4 &  -17.2 &  1.25 & $ 0.63^{+ 0.02}_{- 0.02}$ & $  278^{+   34}_{-   25}$ &   235\tablenotemark{   11} \\
        Cepheus &  109.0 &    7.7 &  1.16 & $ 0.70^{+ 0.03}_{- 0.03}$ & $  709^{+   80}_{-   73}$ &   286\tablenotemark{    3} &         Perseus &  160.7 &  -16.3 &  0.65 & $ 0.69^{+ 0.02}_{- 0.02}$ & $  321^{+   24}_{-   24}$ &   235\tablenotemark{   11} \\
        Cepheus &  114.6 &   16.5 &  0.56 & $ 0.95^{+ 0.02}_{- 0.03}$ & $  366^{+   34}_{-   32}$ &   286\tablenotemark{    3} &         Perseus &  159.9 &  -18.1 &  0.93 & $ 0.79^{+ 0.02}_{- 0.02}$ & $  380^{+   50}_{-   96}$ &   235\tablenotemark{   11} \\
        Cepheus &  113.5 &   15.9 &  0.64 & $ 0.78^{+ 0.03}_{- 0.03}$ & $  389^{+   22}_{-   21}$ &   286\tablenotemark{    3} &         Perseus &  158.5 &  -22.1 &  0.78 & $ 0.66^{+ 0.03}_{- 0.02}$ & $  266^{+   27}_{-   31}$ &   235\tablenotemark{   11} \\
        Cepheus &  116.1 &   20.2 &  0.74 & $ 0.70^{+ 0.02}_{- 0.03}$ & $  321^{+   21}_{-   21}$ &   286\tablenotemark{    3} &         Perseus &  159.3 &  -20.6 &  1.25 & $ 0.54^{+ 0.03}_{- 0.03}$ & $  256^{+   28}_{-   27}$ &   235\tablenotemark{   11} \\
        Cepheus &  111.8 &   20.3 &  1.42 & $ 0.63^{+ 0.02}_{- 0.02}$ & $  365^{+   45}_{-   37}$ &   286\tablenotemark{    3} &         Perseus &  158.6 &  -19.9 &  1.16 & $ 0.61^{+ 0.03}_{- 0.02}$ & $  297^{+   53}_{-   63}$ &   235\tablenotemark{   11} \\
        Cepheus &  112.8 &   16.5 &  0.92 & $ 0.75^{+ 0.03}_{- 0.02}$ & $  396^{+   51}_{-   53}$ &   286\tablenotemark{    3} &         Perseus &  159.7 &  -19.7 &  2.12 & $ 0.54^{+ 0.02}_{- 0.02}$ & $  484^{+  100}_{-  121}$ &   235\tablenotemark{   11} \\
        Cepheus &  108.3 &   17.6 &  1.01 & $ 0.86^{+ 0.03}_{- 0.03}$ & $  372^{+   44}_{-   37}$ &   286\tablenotemark{    3} &         Perseus &  159.9 &  -18.9 &  1.20 & $ 0.66^{+ 0.02}_{- 0.02}$ & $  297^{+   43}_{-   28}$ &   235\tablenotemark{   11} \\
        Cepheus &  111.5 &   12.2 &  1.14 & $ 0.71^{+ 0.01}_{- 0.01}$ & $  883^{+   39}_{-   38}$ &   286\tablenotemark{    3} &         Perseus &  159.4 &  -21.3 &  0.69 & $ 0.52^{+ 0.02}_{- 0.02}$ & $  223^{+   25}_{-   25}$ &   235\tablenotemark{   11} \\
        Cepheus &  110.1 &   17.4 &  1.01 & $ 0.88^{+ 0.02}_{- 0.02}$ & $  317^{+   42}_{-   44}$ &   286\tablenotemark{    3} &         Perseus &  157.8 &  -22.8 &  0.99 & $ 0.70^{+ 0.05}_{- 0.05}$ & $  251^{+   54}_{-   79}$ &   235\tablenotemark{   11} \\
        Cepheus &  104.0 &   14.5 &  0.79 & $ 0.58^{+ 0.01}_{- 0.01}$ & $  673^{+   74}_{-   86}$ &   286\tablenotemark{    3} &         Perseus &  160.4 &  -16.7 &  1.09 & $ 0.46^{+ 0.03}_{- 0.02}$ & $  352^{+   53}_{-   50}$ &   235\tablenotemark{   11} \\
        Cepheus &  103.5 &   13.5 &  0.99 & $ 0.81^{+ 0.01}_{- 0.01}$ & $  372^{+   36}_{-   29}$ &   286\tablenotemark{    3} &         Perseus &  160.8 &  -18.7 &  1.05 & $ 0.82^{+ 0.14}_{- 0.08}$ & $  232^{+   53}_{-   87}$ &   235\tablenotemark{   11} \\
        Cepheus &  112.8 &   20.8 &  0.56 & $ 0.64^{+ 0.04}_{- 0.04}$ & $  401^{+   29}_{-   28}$ &   286\tablenotemark{    3} &         Perseus &  159.1 &  -21.1 &  1.24 & $ 0.51^{+ 0.04}_{- 0.03}$ & $  287^{+   33}_{-   29}$ &   235\tablenotemark{   11} \\
        Cepheus &  111.5 &   20.8 &  0.62 & $ 0.78^{+ 0.02}_{- 0.03}$ & $  247^{+   25}_{-   20}$ &   286\tablenotemark{    3} &         Perseus &  160.8 &  -17.0 &  0.63 & $ 0.80^{+ 0.02}_{- 0.02}$ & $  176^{+   94}_{-   26}$ &   235\tablenotemark{   11} \\
        Cepheus &  115.3 &   17.6 &  0.86 & $ 0.78^{+ 0.02}_{- 0.02}$ & $  390^{+   25}_{-   24}$ &   286\tablenotemark{    3} &         Perseus &  157.7 &  -21.4 &  0.91 & $ 0.75^{+ 0.03}_{- 0.03}$ & $  261^{+   36}_{-   43}$ &   235\tablenotemark{   11} \\
        Cepheus &  107.7 &   12.4 &  0.89 & $ 0.76^{+ 0.01}_{- 0.01}$ & $  957^{+   34}_{-   33}$ &   286\tablenotemark{    3} &         Perseus &  158.2 &  -20.9 &  1.14 & $ 0.65^{+ 0.02}_{- 0.02}$ & $  288^{+   39}_{-   29}$ &   235\tablenotemark{   11} \\
       Hercules &   45.1 &    8.9 &  0.76 & $ 0.60^{+ 0.01}_{- 0.02}$ & $  194^{+    7}_{-    7}$ &                            &         Perseus &  157.5 &  -17.9 &  0.56 & $ 0.86^{+ 0.06}_{- 0.08}$ & $  278^{+   21}_{-   20}$ &   235\tablenotemark{   11} \\
       Hercules &   44.1 &    8.6 &  0.80 & $ 0.57^{+ 0.02}_{- 0.02}$ & $  184^{+    5}_{-    6}$ &                            &         Perseus &  160.0 &  -17.6 &  1.14 & $ 0.69^{+ 0.02}_{- 0.02}$ & $  330^{+   43}_{-   36}$ &   235\tablenotemark{   11} \\
       Hercules &   42.8 &    7.9 &  0.55 & $ 0.52^{+ 0.02}_{- 0.02}$ & $  216^{+    9}_{-    9}$ &                            &         Polaris &  123.5 &   37.9 &  0.16 & $ 0.94^{+ 0.12}_{- 0.14}$ & $  458^{+   66}_{-   75}$ &   100\tablenotemark{   12} \\
        Lacerta &   96.1 &  -10.2 &  0.61 & $ 0.82^{+ 0.03}_{- 0.03}$ & $  517^{+   27}_{-   26}$ &   520\tablenotemark{    4} &         Polaris &  129.5 &   17.3 &  0.47 & $ 0.64^{+ 0.08}_{- 0.04}$ & $  337^{+  166}_{-   44}$ &   100\tablenotemark{   12} \\
        Lacerta &   98.7 &  -14.7 &  0.29 & $ 0.70^{+ 0.09}_{- 0.03}$ & $  322^{+   34}_{-   86}$ &   520\tablenotemark{    4} &         Polaris &  126.3 &   21.2 &  0.44 & $ 0.68^{+ 0.04}_{- 0.03}$ & $  390^{+   34}_{-   34}$ &   100\tablenotemark{   12} \\
        Lacerta &   95.8 &  -11.5 &  0.59 & $ 0.78^{+ 0.03}_{- 0.03}$ & $  509^{+   29}_{-   28}$ &   520\tablenotemark{    4} &         Rosette &  206.8 &   -1.2 &  1.69 & $ 0.61^{+ 0.01}_{- 0.01}$ & $ 1540^{+   69}_{-   67}$ &  1600\tablenotemark{   13} \\
      Maddalena &  217.1 &    0.4 &  2.34 & $ 0.32^{+ 0.01}_{- 0.01}$ & $ 2280^{+   71}_{-   66}$ &  2200\tablenotemark{    5} &         Rosette &  207.8 &   -2.1 &  3.36 & $ 0.50^{+ 0.01}_{- 0.01}$ & $ 1383^{+   85}_{-   64}$ &  1600\tablenotemark{   13} \\
      Maddalena &  216.5 &   -2.5 &  1.89 & $ 0.45^{+ 0.01}_{- 0.01}$ & $ 2222^{+   48}_{-   47}$ &  2200\tablenotemark{    5} &         Rosette &  205.2 &   -2.6 &  1.75 & $ 0.53^{+ 0.01}_{- 0.01}$ & $ 1508^{+   70}_{-   64}$ &  1600\tablenotemark{   13} \\
      Maddalena &  216.8 &   -2.2 &  2.25 & $ 0.50^{+ 0.01}_{- 0.01}$ & $ 2071^{+   59}_{-   55}$ &  2200\tablenotemark{    5} &          Taurus &  171.6 &  -15.8 &  1.07 & $ 0.32^{+ 0.02}_{- 0.01}$ & $  102^{+   25}_{-   32}$ &   140\tablenotemark{   14} \\
      Maddalena &  216.4 &    0.1 &  1.17 & $ 0.54^{+ 0.01}_{- 0.01}$ & $ 2437^{+   69}_{-   71}$ &  2200\tablenotemark{    5} &          Taurus &  175.8 &  -12.9 &  1.54 & $ 0.33^{+ 0.17}_{- 0.03}$ & $  127^{+   26}_{-   34}$ &   140\tablenotemark{   14} \\
        Mon OB1 &  200.4 &    0.8 &  2.20 & $ 0.53^{+ 0.01}_{- 0.01}$ & $  905^{+   61}_{-   55}$ &   913\tablenotemark{    6} &          Taurus &  172.2 &  -14.6 &  1.95 & $ 0.58^{+ 0.01}_{- 0.02}$ & $  128^{+    9}_{-   10}$ &   140\tablenotemark{   14} \\
        Mon OB1 &  201.4 &    1.1 &  2.23 & $ 0.55^{+ 0.01}_{- 0.01}$ & $  887^{+   53}_{-   44}$ &   913\tablenotemark{    6} &          Taurus &  170.2 &  -12.3 &  0.90 & $ 0.53^{+ 0.02}_{- 0.02}$ & $  142^{+   11}_{-   14}$ &   140\tablenotemark{   14} \\
        Mon OB1 &  201.2 &    1.0 &  2.42 & $ 0.54^{+ 0.01}_{- 0.01}$ & $  877^{+   41}_{-   38}$ &   913\tablenotemark{    6} &          Taurus &  166.2 &  -16.6 &  0.66 & $ 0.48^{+ 0.03}_{- 0.03}$ & $  107^{+   18}_{-   20}$ &   140\tablenotemark{   14} \\
         Mon R2 &  219.2 &   -7.7 &  1.21 & $ 0.57^{+ 0.01}_{- 0.01}$ & $ 1018^{+   50}_{-   43}$ &   903\tablenotemark{    7} &          Taurus &  171.4 &  -13.5 &  0.98 & $ 0.52^{+ 0.02}_{- 0.02}$ & $  149^{+    8}_{-    8}$ &   140\tablenotemark{   14} \\
         Mon R2 &  215.3 &  -12.9 &  1.23 & $ 0.65^{+ 0.01}_{- 0.01}$ & $  788^{+   34}_{-   32}$ &   903\tablenotemark{    7} &          Taurus &  173.5 &  -14.2 &  1.98 & $ 0.62^{+ 0.01}_{- 0.01}$ & $  154^{+   14}_{-   21}$ &   140\tablenotemark{   14} \\
         Mon R2 &  219.3 &   -9.5 &  2.35 & $ 0.51^{+ 0.01}_{- 0.01}$ & $ 1026^{+   60}_{-   54}$ &   903\tablenotemark{    7} &         Ursa Ma &  143.4 &   38.5 &  0.35 & $ 0.67^{+ 0.07}_{- 0.07}$ & $  400^{+   33}_{-   27}$ &   110\tablenotemark{   15} \\
         Mon R2 &  220.9 &   -8.3 &  1.12 & $ 0.56^{+ 0.01}_{- 0.01}$ & $ 1052^{+   35}_{-   35}$ &   903\tablenotemark{    7} &         Ursa Ma &  158.5 &   35.2 &  0.21 & $ 1.04^{+ 0.07}_{- 0.06}$ & $  331^{+   29}_{-   25}$ &   110\tablenotemark{   15} \\
         Mon R2 &  213.9 &  -11.9 &  1.32 & $ 0.55^{+ 0.01}_{- 0.01}$ & $  876^{+   42}_{-   41}$ &   903\tablenotemark{    7} &         Ursa Ma &  146.9 &   40.7 &  0.34 & $ 0.57^{+ 0.08}_{- 0.08}$ & $  269^{+   52}_{-   32}$ &   110\tablenotemark{   15} \\
& & & & & & &  Ursa Ma &  153.5 &   36.7 &  0.32 & $ 0.78^{+ 0.07}_{- 0.07}$ & $  353^{+   38}_{-   29}$ &   110\tablenotemark{   15}

\enddata
\tablecomments{
Distances to a selection of molecular clouds.  A number of lines of sight are taken through each cloud and are fit independently.  The Galactic coordinates ($l$, $b$) and the median Planck $E(B-V)$ toward stars used on each sight line are listed.  The best fit normalization $N$ of the Planck dust map in the region is also given, together with its $1\sigma$ uncertainty.  We expect $N = 1$ in diffuse regions, but in these dense and patchy clouds we typically measure $N < 1$, likely due to the finite resolution of emission-based maps.  The distances $D$ are also given, together with uncertainties as 16th and 84th percentiles.  These uncertainties represent the statistical uncertainties alone; an additional 10\% or 15\% overall systematic uncertainty should additionally be included (\textsection\ref{sec:sysunc}).  References represent a small selection of the available literature distances and were used for Figure~\ref{fig:clouddistances}.  References 1--15 refer to \citet{Claria:1974}, \citet{Lada:2009}, \citet{Zdanavicius:2011}, \citet{Kaltcheva:2009}, \citet{Lee:1991}, \citet{Baxter:2009}, \citet{Lombardi:2011}, \citet{Lombardi:2008}, \citet{Menten:2007}, \citet{Murdin:1977}, \citet{Hirota:2008}, \citet{Zagury:1999}, \citet{Blitz:1980}, \citet{Kenyon:1994} and \citet{Penprase:1993}, respectively.
}
\end{deluxetable*}

\tabletypesize{\scriptsize}
\begin{deluxetable*}{cccccc|cccccc}
\tablecaption{Distances to MBM Molecular Clouds
\label{tab:mbmdistances}
}
\tablehead{
\colhead{MBM} & \colhead{$l$ (\degree)} & \colhead{$b$ (\degree)} & \colhead{$E$} & \colhead{$N$} & \colhead{$D$ (pc)} & \colhead{MBM} & \colhead{$l$ (\degree)} & \colhead{$b$ (\degree)} & \colhead{$E$} & \colhead{$N$} & \colhead{$D$ (pc)}
}
\startdata

              1 &  110.2 &  -41.2 &  0.20 & $ 0.81^{+ 0.13}_{- 0.13}$ & $  228^{+   45}_{-   37}$ &             107 &  177.7 &  -20.4 &  0.74 & $ 0.49^{+ 0.02}_{- 0.01}$ & $  197^{+   12}_{-   15}$ \\
              2 &  117.4 &  -52.3 &  0.21 & $ 0.75^{+ 0.09}_{- 0.08}$ & $  206^{+   14}_{-   12}$ &             108 &  178.2 &  -20.3 &  0.72 & $ 0.54^{+ 0.03}_{- 0.03}$ & $  168^{+   19}_{-   21}$ \\
              3 &  131.3 &  -45.7 &  0.23 & $ 0.56^{+ 0.07}_{- 0.06}$ & $  277^{+   22}_{-   26}$ &             109 &  178.9 &  -20.1 &  0.53 & $ 0.54^{+ 0.03}_{- 0.03}$ & $  160^{+   15}_{-   14}$ \\
              4 &  133.5 &  -45.3 &  0.19 & $ 0.70^{+ 0.06}_{- 0.06}$ & $  269^{+   16}_{-   14}$ &             110 &  207.6 &  -22.9 &  0.27 & $ 0.49^{+ 0.03}_{- 0.03}$ & $  313^{+   14}_{-   12}$ \\
              5 &  146.0 &  -49.1 &  0.19 & $ 0.62^{+ 0.09}_{- 0.10}$ & $  187^{+   86}_{-   18}$ &             111 &  208.6 &  -20.2 &  0.77 & $ 0.58^{+ 0.02}_{- 0.02}$ & $  366^{+    9}_{-   11}$ \\
              6 &  145.1 &  -39.3 &  0.23 & $ 0.60^{+ 0.04}_{- 0.03}$ & $  151^{+   16}_{-   66}$ &             113 &  337.7 &   23.0 &  0.39 & $ 0.59^{+ 0.04}_{- 0.04}$ & $  148^{+   11}_{-   13}$ \\
              7 &  150.4 &  -38.1 &  0.27 & $ 0.54^{+ 0.03}_{- 0.02}$ & $  148^{+   13}_{-   11}$ &             115 &  342.3 &   24.1 &  0.35 & $ 0.56^{+ 0.04}_{- 0.04}$ & $  137^{+   12}_{-   16}$ \\
              8 &  151.8 &  -38.7 &  0.27 & $ 0.61^{+ 0.02}_{- 0.02}$ & $  199^{+    9}_{-    9}$ &             116 &  342.7 &   24.5 &  0.36 & $ 0.60^{+ 0.04}_{- 0.04}$ & $  134^{+   10}_{-   11}$ \\
              9 &  156.5 &  -44.7 &  0.18 & $ 0.66^{+ 0.08}_{- 0.08}$ & $  246^{+   42}_{-   29}$ &             117 &  343.0 &   24.1 &  0.33 & $ 0.53^{+ 0.02}_{- 0.02}$ & $  140^{+    9}_{-    9}$ \\
             11 &  158.0 &  -35.1 &  0.20 & $ 0.63^{+ 0.04}_{- 0.03}$ & $  185^{+   21}_{-   20}$ &             118 &  344.0 &   24.8 &  0.29 & $ 0.76^{+ 0.02}_{- 0.03}$ & $   56^{+   21}_{-   17}$ \\
             12 &  159.4 &  -34.3 &  0.55 & $ 0.55^{+ 0.02}_{- 0.02}$ & $  234^{+   11}_{-   10}$ &             119 &  341.6 &   21.4 &  0.19 & $ 0.60^{+ 0.06}_{- 0.05}$ & $  150^{+   26}_{-   32}$ \\
             13 &  161.6 &  -35.9 &  0.25 & $ 0.59^{+ 0.04}_{- 0.04}$ & $  191^{+   11}_{-   12}$ &             120 &  344.2 &   24.2 &  0.38 & $ 0.56^{+ 0.04}_{- 0.02}$ & $   59^{+   70}_{-   23}$ \\
             14 &  162.5 &  -31.9 &  0.23 & $ 0.77^{+ 0.04}_{- 0.04}$ & $  233^{+   11}_{-   10}$ &             121 &  344.8 &   23.9 &  0.36 & $ 0.56^{+ 0.03}_{- 0.02}$ & $  118^{+   11}_{-   13}$ \\
             15 &  191.7 &  -52.3 &  0.20 & $ 0.93^{+ 0.13}_{- 0.13}$ & $  160^{+   47}_{-   64}$ &             122 &  344.8 &   23.9 &  0.36 & $ 0.56^{+ 0.04}_{- 0.02}$ & $  116^{+   11}_{-   19}$ \\
             16 &  170.6 &  -37.3 &  0.77 & $ 0.54^{+ 0.03}_{- 0.03}$ & $  147^{+   10}_{-    9}$ &             123 &  343.3 &   22.1 &  0.27 & $ 0.77^{+ 0.02}_{- 0.03}$ & $  101^{+   12}_{-   19}$ \\
             17 &  167.5 &  -26.6 &  0.30 & $ 0.60^{+ 0.03}_{- 0.02}$ & $  165^{+   16}_{-   14}$ &             124 &  344.0 &   22.7 &  0.29 & $ 0.72^{+ 0.04}_{- 0.05}$ & $   89^{+   16}_{-   16}$ \\
             18 &  189.1 &  -36.0 &  0.51 & $ 0.53^{+ 0.05}_{- 0.04}$ & $  166^{+   18}_{-   17}$ &             125 &  355.5 &   22.5 &  0.58 & $ 0.48^{+ 0.02}_{- 0.03}$ & $  115^{+   16}_{-   14}$ \\
             19 &  186.0 &  -29.9 &  0.27 & $ 0.44^{+ 0.04}_{- 0.03}$ & $  156^{+   18}_{-   18}$ &             126 &  355.5 &   21.1 &  0.82 & $ 0.59^{+ 0.03}_{- 0.03}$ & $  142^{+   12}_{-   17}$ \\
             20 &  210.9 &  -36.6 &  0.44 & $ 0.62^{+ 0.04}_{- 0.04}$ & $  124^{+   11}_{-   14}$ &             127 &  355.4 &   20.9 &  0.97 & $ 0.50^{+ 0.03}_{- 0.03}$ & $  147^{+   12}_{-   12}$ \\
             21 &  208.4 &  -28.4 &  0.22 & $ 0.51^{+ 0.08}_{- 0.09}$ & $  277^{+   23}_{-   22}$ &             128 &  355.6 &   20.6 &  1.10 & $ 0.52^{+ 0.02}_{- 0.02}$ & $  134^{+   11}_{-   11}$ \\
             22 &  208.1 &  -27.5 &  0.17 & $ 0.84^{+ 0.10}_{- 0.11}$ & $  238^{+   27}_{-   22}$ &             129 &  356.2 &   20.8 &  1.02 & $ 0.54^{+ 0.03}_{- 0.03}$ & $  141^{+   11}_{-   11}$ \\
             23 &  171.8 &   26.7 &  0.18 & $ 0.74^{+ 0.06}_{- 0.06}$ & $  305^{+   22}_{-   22}$ &             130 &  356.8 &   20.3 &  0.94 & $ 0.54^{+ 0.03}_{- 0.03}$ & $  109^{+   10}_{-   13}$ \\
             24 &  172.3 &   27.0 &  0.18 & $ 0.69^{+ 0.08}_{- 0.07}$ & $  279^{+   27}_{-   23}$ &             131 &  359.2 &   21.8 &  0.68 & $ 0.62^{+ 0.03}_{- 0.03}$ & $  106^{+   11}_{-   11}$ \\
             25 &  173.8 &   31.5 &  0.16 & $ 0.76^{+ 0.10}_{- 0.12}$ & $  297^{+   25}_{-   27}$ &             132 &    0.8 &   22.6 &  0.67 & $ 0.53^{+ 0.04}_{- 0.03}$ & $  155^{+    9}_{-   10}$ \\
             27 &  141.3 &   34.5 &  0.18 & $ 0.89^{+ 0.08}_{- 0.08}$ & $  359^{+   23}_{-   21}$ &             133 &  359.2 &   21.4 &  0.65 & $ 0.63^{+ 0.02}_{- 0.02}$ & $   98^{+   12}_{-   11}$ \\
             28 &  141.4 &   35.2 &  0.19 & $ 0.86^{+ 0.07}_{- 0.07}$ & $  370^{+   22}_{-   20}$ &             134 &    0.1 &   21.8 &  0.77 & $ 0.55^{+ 0.03}_{- 0.03}$ & $  121^{+   14}_{-   28}$ \\
             29 &  142.3 &   36.2 &  0.20 & $ 0.69^{+ 0.13}_{- 0.13}$ & $  376^{+   76}_{-   60}$ &             135 &    2.7 &   22.0 &  0.62 & $ 0.53^{+ 0.01}_{- 0.01}$ & $  180^{+   11}_{-   10}$ \\
             30 &  142.2 &   38.2 &  0.22 & $ 0.80^{+ 0.05}_{- 0.05}$ & $  352^{+   10}_{-   11}$ &             136 &    1.3 &   21.0 &  0.61 & $ 0.64^{+ 0.04}_{- 0.04}$ & $  120^{+   12}_{-   12}$ \\
             31 &  146.4 &   39.6 &  0.18 & $ 0.88^{+ 0.09}_{- 0.09}$ & $  325^{+   27}_{-   26}$ &             137 &    4.5 &   22.9 &  0.64 & $ 0.57^{+ 0.03}_{- 0.03}$ & $  146^{+   11}_{-   16}$ \\
             32 &  147.2 &   40.7 &  0.22 & $ 0.89^{+ 0.07}_{- 0.07}$ & $  259^{+   14}_{-   15}$ &             138 &    3.1 &   21.8 &  0.60 & $ 0.54^{+ 0.01}_{- 0.01}$ & $  186^{+    8}_{-    8}$ \\
             33 &  359.0 &   36.8 &  0.19 & $ 0.63^{+ 0.02}_{- 0.03}$ & $   88^{+   18}_{-   21}$ &             139 &    7.7 &   24.9 &  0.37 & $ 0.68^{+ 0.08}_{- 0.07}$ & $  112^{+   10}_{-   26}$ \\
             34 &    2.3 &   35.7 &  0.19 & $ 0.61^{+ 0.07}_{- 0.08}$ & $  110^{+   27}_{-   34}$ &             140 &    3.2 &   21.7 &  0.61 & $ 0.54^{+ 0.01}_{- 0.01}$ & $  186^{+    8}_{-    9}$ \\
             35 &    6.6 &   38.1 &  0.22 & $ 0.74^{+ 0.04}_{- 0.07}$ & $   89^{+   17}_{-   25}$ &             141 &    4.8 &   22.6 &  0.62 & $ 0.60^{+ 0.03}_{- 0.03}$ & $  127^{+   13}_{-   14}$ \\
             36 &    4.2 &   35.8 &  0.42 & $ 0.53^{+ 0.03}_{- 0.03}$ & $  105^{+    7}_{-    7}$ &             142 &    3.6 &   21.0 &  0.58 & $ 0.65^{+ 0.04}_{- 0.05}$ & $  133^{+   14}_{-   13}$ \\
             37 &    6.1 &   36.8 &  0.42 & $ 0.50^{+ 0.02}_{- 0.01}$ & $  121^{+   10}_{-   16}$ &             143 &    6.0 &   20.2 &  0.62 & $ 0.53^{+ 0.03}_{- 0.02}$ & $  131^{+    8}_{-    6}$ \\
             38 &    8.2 &   36.3 &  0.23 & $ 0.68^{+ 0.06}_{- 0.06}$ & $   77^{+   24}_{-   24}$ &             144 &    6.6 &   20.6 &  0.65 & $ 0.50^{+ 0.03}_{- 0.02}$ & $  128^{+    9}_{-    9}$ \\
             39 &   11.4 &   36.3 &  0.21 & $ 0.65^{+ 0.05}_{- 0.04}$ & $   94^{+   15}_{-   11}$ &             145 &    8.5 &   21.9 &  0.58 & $ 0.51^{+ 0.02}_{- 0.01}$ & $  152^{+   19}_{-   25}$ \\
             40 &   37.6 &   44.7 &  0.20 & $ 0.80^{+ 0.05}_{- 0.06}$ & $   64^{+   21}_{-   25}$ &             146 &    8.8 &   22.0 &  0.50 & $ 0.53^{+ 0.02}_{- 0.01}$ & $  179^{+   11}_{-   12}$ \\
             45 &    9.8 &  -28.0 &  0.20 & $ 0.63^{+ 0.10}_{- 0.09}$ & $  131^{+   21}_{-   29}$ &             147 &    5.9 &   20.1 &  0.60 & $ 0.55^{+ 0.03}_{- 0.02}$ & $  130^{+    8}_{-    7}$ \\
             46 &   40.5 &  -35.5 &  0.26 & $ 0.85^{+ 0.05}_{- 0.05}$ & $  490^{+   25}_{-   23}$ &             148 &    7.5 &   21.1 &  0.77 & $ 0.65^{+ 0.03}_{- 0.03}$ & $  116^{+   10}_{-   10}$ \\
             47 &   41.0 &  -35.9 &  0.29 & $ 0.88^{+ 0.04}_{- 0.04}$ & $  475^{+   25}_{-   21}$ &             149 &    7.9 &   20.3 &  0.81 & $ 0.67^{+ 0.03}_{- 0.03}$ & $  114^{+   13}_{-   11}$ \\
             49 &   64.5 &  -26.5 &  0.18 & $ 0.60^{+ 0.09}_{- 0.09}$ & $  204^{+   39}_{-   33}$ &             150 &    9.6 &   21.3 &  0.52 & $ 0.59^{+ 0.03}_{- 0.03}$ & $  139^{+   14}_{-   12}$ \\
             50 &   70.0 &  -31.2 &  0.18 & $ 0.67^{+ 0.09}_{- 0.08}$ & $   99^{+   43}_{-   45}$ &             151 &   21.5 &   20.9 &  0.41 & $ 0.62^{+ 0.03}_{- 0.03}$ & $  122^{+    8}_{-    8}$ \\
             53 &   94.0 &  -34.1 &  0.23 & $ 0.71^{+ 0.05}_{- 0.04}$ & $  253^{+   10}_{-   11}$ &             156 &  101.7 &   22.8 &  0.19 & $ 0.75^{+ 0.04}_{- 0.04}$ & $  313^{+   12}_{-   10}$ \\
             54 &   91.6 &  -38.1 &  0.20 & $ 0.67^{+ 0.06}_{- 0.05}$ & $  231^{+   11}_{-   12}$ &             157 &  103.2 &   22.7 &  0.20 & $ 0.70^{+ 0.04}_{- 0.04}$ & $  325^{+   11}_{-    9}$ \\
             55 &   89.2 &  -40.9 &  0.27 & $ 0.70^{+ 0.04}_{- 0.04}$ & $  206^{+    8}_{-    6}$ &             158 &   27.2 &  -20.7 &  0.23 & $ 0.54^{+ 0.02}_{- 0.02}$ & $  142^{+    9}_{-   10}$ \\
             56 &  103.1 &  -26.1 &  0.23 & $ 0.74^{+ 0.05}_{- 0.05}$ & $  227^{+   17}_{-   17}$ &             159 &   27.4 &  -21.1 &  0.23 & $ 0.57^{+ 0.01}_{- 0.01}$ & $  143^{+    8}_{-   10}$ \\
             57 &    5.1 &   30.8 &  0.23 & $ 0.58^{+ 0.06}_{- 0.04}$ & $   88^{+   25}_{-   39}$ &             161 &  114.7 &   22.5 &  0.23 & $ 0.71^{+ 0.04}_{- 0.04}$ & $  308^{+    8}_{-    8}$ \\
            101 &  158.2 &  -21.4 &  1.14 & $ 0.49^{+ 0.02}_{- 0.01}$ & $  283^{+   11}_{-   10}$ &             162 &  111.7 &   20.1 &  0.66 & $ 0.51^{+ 0.02}_{- 0.02}$ & $  304^{+    8}_{-    9}$ \\
            102 &  158.6 &  -21.2 &  1.46 & $ 0.46^{+ 0.01}_{- 0.01}$ & $  275^{+    9}_{-    9}$ &             163 &  115.8 &   20.2 &  0.46 & $ 0.69^{+ 0.02}_{- 0.02}$ & $  293^{+    6}_{-    7}$ \\
            103 &  158.9 &  -21.6 &  1.00 & $ 0.47^{+ 0.02}_{- 0.02}$ & $  269^{+   10}_{-    9}$ &             164 &  116.2 &   20.4 &  0.41 & $ 0.69^{+ 0.02}_{- 0.02}$ & $  294^{+    6}_{-    6}$ \\
            104 &  158.4 &  -20.4 &  1.52 & $ 0.48^{+ 0.01}_{- 0.01}$ & $  262^{+    9}_{-    9}$ &             165 &  116.2 &   20.3 &  0.44 & $ 0.68^{+ 0.02}_{- 0.02}$ & $  291^{+    7}_{-    7}$ \\
            105 &  169.5 &  -20.1 &  0.37 & $ 0.46^{+ 0.01}_{- 0.01}$ & $  139^{+    8}_{-    9}$ &             166 &  117.4 &   21.5 &  0.26 & $ 0.65^{+ 0.03}_{- 0.03}$ & $  302^{+   10}_{-   10}$ \\
            106 &  176.3 &  -20.8 &  0.55 & $ 0.55^{+ 0.01}_{- 0.01}$ & $  190^{+   12}_{-   16}$ & & & & & &

\enddata
\tablecomments{
Distances to the high-latitude molecular clouds of \citet{Magnani:1985}.  The Galactic coordinates ($l$, $b$) and the median Planck $E(B-V)$ toward stars used on each sight line are listed.  The best fit normalization $N$ of the Planck dust map in the region is also given, together with $1\sigma$ uncertainty; we expect $N = 1$ in diffuse regions, but in these dense and patchy clouds we typically measure $N < 1$, likely due to the difference between how emission-based maps and stellar-based maps are constructed (see \textsection\ref{sec:discussion}).  Our distance estimates $D$ are also given.  For distance uncertainties, 16th and 84th percentiles are given.  An additional 10\% or 15\% overall systematic uncertainty should be additionally included to account for limitations of our technique (\textsection\ref{sec:sysunc}).  For a few clouds in the MBM catalog the total extinction in the region is too low for a reliable distance fit---we have excluded those clouds without stars with Planck $E(B-V) > 0.15$.  A small number of MBM clouds with $\delta < -30\degree$ that do not fall in the \PS\ footprint are also excluded.
}
\end{deluxetable*}

\figsetstart
\figsetnum{4}
\figsettitle{Supplementary Cloud Distance Determination Figures}

\figsetgrpstart
\figsetgrpnum{4.0}
\figsetgrptitle{Distance to Aquila S}
\figsetplot{figset/AquilaS_39.3_-16.8}
\figsetgrpnote{Determination of the distance to Aquila S, using a sightline through ($l$, $b$) = (39.3\degree, -16.8\degree).  See Figure~\ref{fig:distancefits} caption for details.}
\figsetgrpend

\figsetgrpstart
\figsetgrpnum{4.65}
\figsetgrptitle{Distance to Orion}
\figsetplot{figset/Orion_208.4_-19.6}
\figsetgrpnote{Determination of the distance to Orion, using a sightline through ($l$, $b$) = (208.4\degree, -19.6\degree).  See Figure~\ref{fig:distancefits} caption for details.}
\figsetgrpend

\figsetgrpstart
\figsetgrpnum{4.117}
\figsetgrptitle{Distance to Taurus}
\figsetplot{figset/Taurus_172.2_-14.6}
\figsetgrpnote{Determination of the distance to Taurus, using a sightline through ($l$, $b$) = (172.2\degree, -14.6\degree).  See Figure~\ref{fig:distancefits} caption for details.}
\figsetgrpend

\figsetgrpstart
\figsetgrpnum{4.135}
\figsetgrptitle{Distance to MBM 12}
\figsetplot{figset/MBM12_159.4_-34.3}
\figsetgrpnote{Determination of the distance to MBM 12, using a sightline through ($l$, $b$) = (159.4\degree, -34.3\degree).  See Figure~\ref{fig:distancefits} caption for details.}
\figsetgrpend

\figsetend

\begin{figure*}[htb]
\figurenum{4}
\dfplot{\detokenize{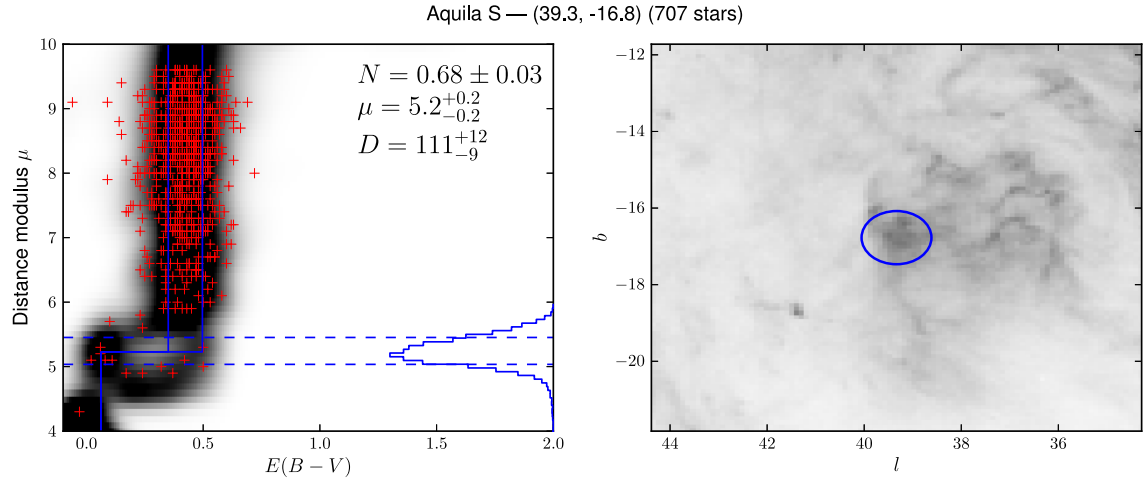}}
\figcaption{
Determination of the distance to Aquila S, using a sightline through ($l$, $b$) = (39.3\degree, -16.8\degree).  See Figure~\ref{fig:distancefits} caption for details.  The ApJ online version of the paper and our web site include analogous Figures for each sight line.
}
\end{figure*}

\end{document}